\documentclass{aa}
\pdfoutput=1
\usepackage[varg]{txfonts}
\usepackage{natbib}
\usepackage{caption}
\usepackage{subcaption}
\usepackage{chemformula}
\usepackage{pdflscape}
\usepackage{afterpage}
\usepackage{xcolor}
\usepackage{booktabs}
\usepackage{amsmath,bm}
\usepackage{multirow}
\usepackage{lscape}
\usepackage{rotating}

\usepackage{color}
\usepackage{natbib,twoopt}
\usepackage[hyphenbreaks]{breakurl}
\usepackage[breaklinks]{hyperref}      
\bibpunct{(}{)}{;}{a}{}{,}             
\definecolor{cobalt}{rgb}{0.06, 0.2, 0.65}
\hypersetup{
  colorlinks,
  citecolor=cobalt,
  linkcolor=[rgb]{0.8, 0.2, 1.0},
  urlcolor=cobalt,
}
\makeatletter
  \newcommandtwoopt{\citeads}[3][][]{\href{http://adsabs.harvard.edu/abs/#3}%
    {\def\hyper@linkstart##1##2{}%
     \let\hyper@linkend\@empty\citealp[#1][#2]{#3}}}
  \newcommandtwoopt{\citepads}[3][][]{\href{http://adsabs.harvard.edu/abs/#3}%
    {\def\hyper@linkstart##1##2{}%
     \let\hyper@linkend\@empty\citep[#1][#2]{#3}}}
  \newcommandtwoopt{\citetads}[3][][]{\href{http://adsabs.harvard.edu/abs/#3}%
    {\def\hyper@linkstart##1##2{}%
     \let\hyper@linkend\@empty\citet[#1][#2]{#3}}}
  \newcommandtwoopt{\citeyearads}[3][][]%
    {\href{http://adsabs.harvard.edu/abs/#3}
    {\def\hyper@linkstart##1##2{}%
     \let\hyper@linkend\@empty\citeyear[#1][#2]{#3}}}
   \renewcommand*\aa@pageof{, page \thepage{} of \pageref*{LastPage}}
\makeatother

\usepackage[separate-uncertainty = true,multi-part-units=single,group-separator={,}, group-digits = integer, group-four-digits = true,per-mode=reciprocal]{siunitx}

\newcommand{\bibi}{WASP-189\,b\xspace}
\newcommand{\numpm}[3]{$#1^{#2}_{#3}$}
\newcommand{\hoeijmakers}{Hoeijmakers et al. (in prep)\xspace}
\newcommand{\lam}{Lam et al. (in prep)\xspace}

\definecolor{lightblue}{rgb}{0.2, 0.6, 1}
\newcommand{\citesoftware}[2]{{\href{#2}{\texttt{\color{cobalt} #1}}}\xspace}
\newcommand{\StarRotator}{\citesoftware{StarRotator}{https://github.com/Hoeijmakers/StarRotator}}

\newcommand{\alldetections}{\ch{H}, \ch{Na}, \ch{Mg}, \ch{Ca}, \ch{Ca+}, \ch{Ti}, \ch{Ti+}, \ch{TiO}, \ch{V}, \ch{Cr}, \ch{Mn}, \ch{Fe}, \ch{Fe+}, \ch{Ni}, \ch{Sr}, \ch{Sr+}, and \ch{Ba+}\xspace}

\begin{document}
\title{An atlas of resolved spectral features in the transmission spectrum of \bibi with MAROON-X}
\titlerunning{A spectral atlas of \bibi with MAROON-X}
\authorrunning{Prinoth et al. (2024)}

\author{                                           
 B.\,Prinoth\inst{1,2}                             
\and H.\,J.\,Hoeijmakers\inst{2}                   
\and B.\,M.\,Morris\inst{3}                        
\and M. Lam\inst{2}
\and D. Kitzmann\inst{4}
\and E. Sedaghati\inst{1}                        
\and J.\,V.\,Seidel\inst{1}                      
\and E.\,K.\,H.\,Lee\inst{4}
\and B.\,Thorsbro\inst{5,2}                      
\and N.\,W.\,Borsato\inst{6,2}
\and Y.\,C.\,Damasceno\inst{7,8}
\and S.\,Pelletier\inst{9}                       
\and A.\,Seifahrt\inst{10}  
}

\offprints{Bibiana Prinoth, \\ \email{bibiana.prinoth@fysik.lu.se}}

\institute{
European Southern Observatory, Alonso de C\'ordova 3107, Vitacura, Regi\'on Metropolitana, Chile 
\and Lund Observatory, Division of Astrophysics, Department of Physics, Lund University, Box 43, 221 00 Lund, Sweden  
\and Space Telescope Science Institute, Baltimore, MD 21218, USA 
\and University of Bern, Physics Institute, Division of Space Research \& Planetary Sciences, Gesellschaftsstr. 6, 3012, Bern, Switzerland 
\and Observatoire de la C\^ote d'Azur, CNRS UMR 7293, BP4229, Laboratoire Lagrange, F-06304 Nice Cedex 4, France 
\and School of Mathematical and Physical Sciences, Macquarie University, Sydney, NSW 2109, Australia 
\and Instituto de Astrof\'isica e Ci\^encias do Espa\c{c}o, Universidade do Porto, CAUP, Rua das Estrelas, 4150-762 Porto, Portugal 
\and Departamento de F\'isica e Astronomia, Faculdade de Ci\^encias, Universidade do Porto, Rua do Campo Alegre, 4169-007 Porto, Portugal 
\and Department of Physics and Trottier Institute for Research on Exoplanets, University of Montreal, Montreal, QC, Canada 
\and Department of Astronomy \& Astrophysics, The University of Chicago, Chicago, IL, USA 
}

\date{Received}

\abstract{Exoplanets in the ultra-hot Jupiter regime provide an excellent laboratory for testing the impact of stellar irradiation on the dynamics and chemical composition of gas giant atmospheres. In this study, we observed two transits of the ultra-hot Jupiter \bibi with MAROON-X/Gemini-North to probe its high-altitude atmospheric layers, using strong absorption lines. We derived posterior probability distributions for the planetary and stellar parameters by calculating the stellar spectrum behind the planet at every orbital phase during the transit. This was used to correct the Rossiter-McLaughlin imprint on the transmission spectra. Using differential transmission spectroscopy, we detect strong absorption lines of \ch{Ca+}, \ch{Ba+}, \ch{Na}, \ch{H$\alpha$}, \ch{Mg}, \ch{Fe}, and \ch{Fe+}, providing an unprecedented and detailed view of the atmospheric chemical composition. \ch{Ca+} absorption is particularly well suited for analysis through time-resolved narrow-band spectroscopy, owing to its transition lines formed in high-altitude layers.  The spectral absorption lines show no significant blueshifts that would indicate high-altitude day-to-night winds, and further analysis is needed to investigate the implications for atmospheric dynamics. These high signal-to-noise observations provide a benchmark data set for testing high-resolution retrievals and the assumptions of atmospheric models. We also simulate observations of \bibi with ANDES/ELT, and show that ANDES will be highly sensitive to the individual absorption lines of a myriad of elements and molecules, including TiO and CO.}
\maketitle

\section{Introduction}
\label{sec:Introduction}
\begin{table*}
        \caption{Overview of observations.}\vspace{-1em}
        \begin{center}
                \begin{tabular}{llllll}
                        \toprule
                        Date   & $\#$ Spectra$^a$  & t$_{\rm exp}$ [s] $^{b}$ & Airmass $^{c}$ & Avg. S/N & Min./Max. seeing $^{d}$ \\
                        \midrule
                        2022-04-03   & 57 (38/19)  & b: 200, r: 160  & 3.07$-$1.09$-$1.22 & b: 168 r: 202 & 0.49 / 1.87 \\
                        2022-06-02   & 57 (40/17)  & b: 200, r: 160  & 1.52$-$1.09$-$1.72 & b: 169 r: 203 & 0.29 / 1.03 \\
                        \bottomrule
                \end{tabular}
        \end{center}
        \vspace{-1em}
        \textit{Note:} $^{a}$ In parentheses, in-transit and out-of-transit spectra, respectively. $^{b}$ Exposure times differ for the blue and red arms of MAROON-X. $^{c}$ Airmass at the start and end of the observation, as well as minimum airmass at the highest altitude of the target. $^{d}$\,The seeing is taken from the Maunakea Weather Center\footnote{\url{http://mkwc.ifa.hawaii.edu/current/seeing/}} for the two nights of observations.
        \label{tab:observation_log}
\end{table*}
The primary transit of an exoplanet, which is when it passes in front of the disc of its host star, presents a unique opportunity to study its atmosphere through the imprint it leaves on the traversing stellar radiation. This imprint is caused by absorption at specific wavelengths that correspond to the species present in the upper atmosphere. The components of this absorption depend on the environmental conditions they reside in, namely the pressure and temperature structure. One approach to extracting these aforementioned signatures involves performing differential spectroscopy, whereby out-of-transit spectra are used to remove the star from the in-transit observations, thus leaving only the planetary signal in the residual spectra. Each of these planetary spectra is Doppler-shifted from the rest frame by the radial velocity of the planet at the time of each observation. This method was successfully applied by \citet{wyttenbach_spectrally_2015} to extract the planetary spectrum of HD\,189733\,b and confirm the presence of neutral sodium using the HARPS spectrograph \citep{mayor_setting_2003}, by placing the residual spectra in the planetary rest frame and combining observations of three separate transit events. Subsequently, neutral sodium has been detected in a number of close-in giant planet atmospheres through this methodology \citep[e.g.][]{casasayas-barris_na_2018, jensen_hydrogen_2018, seidel_hot_2019, chen_HARPS_2020, tabernero_espresso_2021, borsa_atmospheric_2021, mounzer_hot_2022,seidel_prinoth_detection_2023}. Sodium is a powerful diagnostic for studying the upper atmosphere of exoplanets, in particular using its Frauenhofer D-lines. These resonant lines probe high up in the atmosphere and even relatively low sodium abundances provide strong absorption features owing to the large atomic cross-sections. Evidently, sodium is not the only species where this so-called narrow-band spectroscopy can be used to study the higher altitudes of exoplanetary atmospheres. Other significant transition lines at optical to near-infrared wavelengths also probe these regions and have readily been detected, some of which include the \ch{He} triplet \citep{allart_He_2018, Kirk_He_2020, kirk_He_2022, bello-arufe_exoplanet_2022,orell_He_2023}, the \ch{Ca+} triplet \citep{casasayas-barris_carmenes_2021, bello-arufe_exoplanet_2022}, or the H$\alpha$ line of the Balmer series \citep{chen_HARPS_2020, bello-arufe_exoplanet_2022,seidel_prinoth_detection_2023}. Furthermore, signatures of \ch{Mg} and \ch{Li} \citep{borsa_atmospheric_2021}, as well as Paschen-$\alpha$ \citep{sanchez-lopez_detection_2022}, have been detected, probing deeper layers in atmospheres.

Even for the routinely detected sodium, one needs to shift the in-transit residual spectra (planetary spectra) to the planetary rest frame and combine them to boost the signal-to-noise ratio, thus pushing the planetary signal above the noise floor. However, doing so removes any temporal and spatial information inherent in the observations. Furthermore, combining multiple transit events observed over a time span sometimes lasting years, any intrinsic variability in the atmosphere \citep{2012Lecavelier} is averaged out, reducing the amount of information that is extracted from the transmission spectrum.

A complementary approach is the use of the cross-correlation technique \citep{snellen_orbital_2010}, whereby signals from a multitude of absorption lines are combined by placing them in velocity space. This enables some of the aforementioned limitations to be overcome either directly from observations \citep{borsa_gaps_2019, ehrenreich_nightside_2020, kesseli_confirmation_2021, kesseli_atomic_2022, pelletier_vanadium_2023,prinoth_time-resolved_2023} or via retrieval techniques \citep{gandhi_spatially_2022, gandhi_retrieval_2023}, eschewing the need to integrate over the transit. In this approach, one preserves temporal and spatial information embedded in each in-transit spectrum, allowing for a more detailed view of the atmosphere.

To achieve a similar paradigm shift for narrow-band transmission spectroscopy towards time-resolved studies, a detectable signal in each spectrum is required, achieved either by employing superior photon-collecting power \citep{seidel_hot_2022} or by observing extremely hot targets with inherently large signatures \citep{pino_neutral_2020}. 

In this work, we contribute to the ongoing revolution of narrow-band transmission spectroscopy by providing time-resolved spectra of the atmosphere of the ultra-hot Jupiter exoplanet \bibi. It has a radius of 1.619\,$R_{\rm Jup}$ and a mass of 1.99\,$M_{\rm Jup}$, orbiting its bright ($V=6.6$) A star on a 2.72\,day polar orbit, with an estimated equilibrium temperature of $\sim$\,2600\,K \citep{anderson_wasp-189b_2018}. Previous studies have demonstrated the rich atmospheric inventory of this planet's transmission spectrum using the cross-correlation technique, revealing detections of \alldetections \citep{stangret_high-resolution_2022,prinoth_titanium_2022,prinoth_time-resolved_2023}, as well as time-resolved signals in the cross-correlation space \citep{prinoth_time-resolved_2023}. With the observations presented here, we demonstrate that the high-resolution, stabilised spectra needed for time-resolved narrow-band transmission spectroscopy can be obtained with MAROON-X on the 8\,m class Gemini-North, thus opening up a new channel of study for the community.

The \bibi system provides an observational sweet spot, where the brightness and geometry of the system allow high signal-to-noise ratio spectra to be obtained at relatively short exposure times, minimising the effect of smearing, which is crucial to preserve the line shapes \citep{boldt-christmas_optimising_2023}. Short exposure times make it possible to resolve single lines while keeping the information about their temporal variations, such for example the \ch{Ca+} triplet with a single transit event. This allows for the modelling of the planetary absorption combined with the residual signal caused by the planet crossing the stellar disc (the Rossiter-McLaughlin effect). In short, no stacking in time is required, which enables time-resolved studies of the planetary absorption. Additionally, we analyse the regions of the \ch{Na} Frauenhofer D-lines, the \ch{H$\alpha$} line, the \ch{Mg} triplet, the \ch{Ca+} infrared triplet, and multiple \ch{Ba+} lines, as well as several stronger \ch{Fe} and \ch{Fe+} lines to search for planetary absorption. Multiple absorption lines for different chemical species at high significance invite further investigation, especially within the framework of atmospheric retrievals for both dynamics and composition at high spectral resolution.

This manuscript is structured as follows. In Section \ref{sec:observations}, we describe the observations and data reduction. Section \ref{sec:methods} covers the methodology, including corrections for telluric absorption, outliers, and velocities. It further includes the methods of extracting the transmission spectrum and introduces our treatment of time-resolved and classical narrow-band spectroscopy. The results are presented and discussed in Section \ref{sec:results_discussion} and concluded in Section \ref{sec:conclusion}.

\begin{figure*}[t!]
    \centering
    \includegraphics[width=\linewidth]{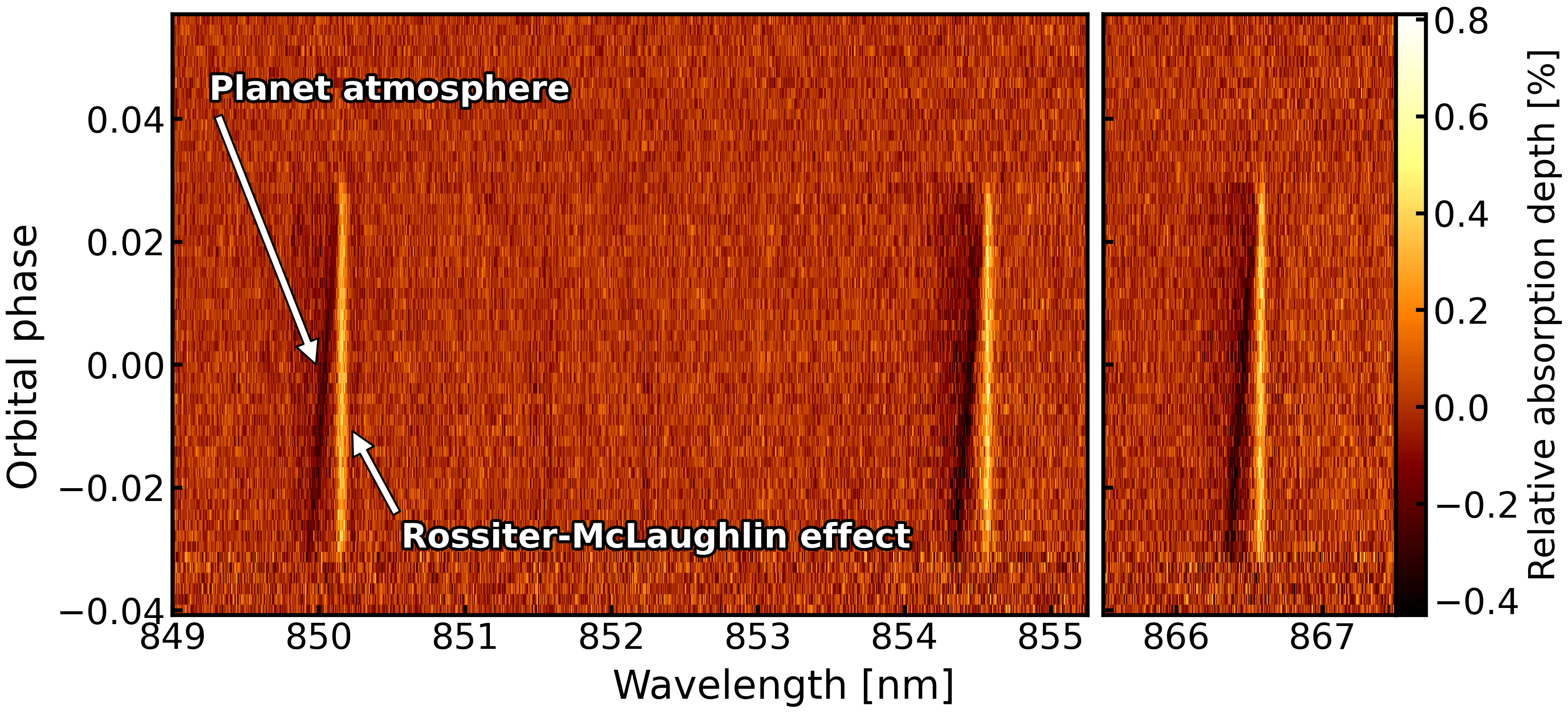}
    \caption{Spectral time series at the wavelength of interest for the \ch{Ca+} triplet around 850 -- 867 nm. The time series of the two observation nights were stacked after correcting for telluric contamination, and normalisation of the spectra to a common flux level, as well as outlier rejection with subsequent integration over the outlier pixels. The spectral time series is shown in the rest frame of the star. The vertical bright emission feature originates from the planet crossing the stellar disc over the course of the transit (Rossiter-McLaughlin effect or Doppler shadow). The dark absorption feature is caused by the atmosphere of the planet. In addition, there is a fainter, wider dark absorption feature, barely noticeable by eye, which is also part of the residual of the planet crossing the stellar disc (see Fig\,.\ref{fig:stepbystep}). This spectral time series is a rare case in which you can see the planetary absorption without stacking in the planetary rest frame thanks to the high signal-to-noise ratio achieved during these MAROON-X observations.}
    \label{fig:order_22_23}
\end{figure*}

\section{Observations and data reduction}
\label{sec:observations}

We observed two transit time series of \bibi on April 3, 2022 and June 2, 2022 (programme ID: GN-2022A-FT-208, PI: Pelletier) with the MAROON-X high-resolution, cross-dispersed, echelle optical spectrograph mounted on the 8.1-m Gemini-North telescope in Hawaii \citep{seifahrt_maroon-x_2018,seifahrt_-sky_2020}. It covers the wavelength range from \num{490} to \SI{920}{nm} within a blue and a red arm, at a spectral resolving power of $R \approx\num{85000}$. The observations cover the full transit as well as baseline exposures before and after the transit. The observations were reduced using the dedicated pipeline \citep{seifahrt_-sky_2020} (see also \cite{prinoth_time-resolved_2023} for details on these data sets). 

During both transit observations, 57 spectra were taken, whereas 38 and 40 spectra of those were in transit for April 3, 2022 and June 2, 2022, respectively. Due to the different readout times of the detectors, the exposure times in the blue and red arms were \SI{200}{\s} and \SI{160}{\s}. More information on the two sets of observations is provided in Table\,\ref{tab:observation_log}. 

We also determined the systemic velocity by computing the stellar cross-correlation functions for the out-of-transit exposures with a PHOENIX \citep{husser_new_2013} template of the star, following \citet{zhang_detection_2022}. By fitting a rotationally broadened model \citep{gray_observation_2008} to the averaged out-of-transit cross-correlation functions, we determined the systemic velocity to be \SI{-22.71 \pm 0.67}{\km\per\second} and \SI{-23.65 \pm 0.67}{\km\per\second} for the two data sets, consistent with the value determined by \citet{anderson_wasp-189b_2018} ($\SI{-24.45 \pm 0.01}{\km\per\second}$). We adopted this latter value for the rest of the study for comparability with previous work.

\section{Methods}
\label{sec:methods}

\subsection{Preparatory corrections}

The reduced spectra were corrected for telluric contamination with \texttt{molecfit} \citep[v1.5.9, ][]{smette_molecfit_2015,kausch_molecfit_2015}. For each exposure in the time series, regions with strong \ch{H2O} and \ch{O2} absorption lines were selected to compute the telluric model, accounting for the changing weather conditions, airmass, and seeing. These telluric models were then interpolated onto the same wavelength grid as the data and divided out. Residual telluric contamination was later manually masked where needed. The individual spectra were Doppler-shifted to the rest frame of the star, accounting for the Earth's velocity around the barycentre of the Solar System, $\bm{v}_{\rm BERV}$, the radial velocity of the star caused by the orbiting planet, $\bm{v}_{\ast, \rm RV} $, and the velocity of the planetary system, $v_{\rm sys}$, using
\begin{equation}
    \bm{v}_{\rm corr} = \bm{v}_{\rm BERV} - \bm{v}_{\ast, \rm RV} - v_{\rm sys}.
\end{equation}

\noindent This yields a Doppler correction for the wavelength of the form
\begin{equation}
    \bm{\lambda}_{\rm corr} = \left(1 + \frac{\bm{v}_{\rm corr}}{c}\right) \bm{\lambda},
    \label{eq:wl_shift}
\end{equation}
\noindent where $c$ is the speed of light and $\bm{\lambda}$ is the wavelength as observed by the spectrograph. Following \citet{hoeijmakers_high-resolution_2020}, we corrected for outliers by applying an order-by-order sigma clipping algorithm. We calculated a running median absolute deviation over sub-bands of the time series with a width of 40 pixels and rejected $5\sigma$-outliers. 
The spectra were colour-corrected using a polynomial of degree 2 (see \citet{hoeijmakers_hot_2020}), which accounts for colour-dependent variations in the illumination. We did not use the blaze-corrected spectra, as the blaze was removed during the division of the out-of-transit baseline. Using fibre B, which was on sky during the observations, we masked atmospheric emission features.

\subsection{Transmission spectra}

We obtained the transmission spectra of the planet by dividing the normalised in-transit spectra by the normalised master out-of-transit spectrum; in other words, by undertaking differential transmission spectroscopy, following \citet{wyttenbach_spectrally_2015}. This master out-of-transit spectrum is an averaged spectrum of all out-of-transit exposures. 
Through this division, one obtains the normalised transmission spectra in the rest frame of the star. Typically, at this stage, the planetary signature is not visible in the time series due to the relatively weak absorption and low signal-to-noise ratio. However, this is not the case for these data sets. 
For instance, for the \ch{Ca+} infrared triplet, the planetary absorption and the residual of the stellar spectrum obscured by the planet during transit (the Rossiter-McLaughlin effect) are astonishingly visible (see Fig. \ref{fig:order_22_23}), making these data sets a true benchmark. 
A high signal-to-noise ratio is thus key to observing single, temporally resolved absorption lines.
In the rest frame of the star (see Fig.\,\ref{fig:order_22_23}), the planetary atmosphere signal traces the radial velocity of the planet during transit, described through a sinusoidal dependence on the orbital phase, $\phi$. This radial velocity change introduces a shift in the wavelength of the planetary absorption described by
\begin{equation}
     \bm{\lambda}_{\rm p}(\bm{\phi}) =  \Delta\lambda_{\rm max} \sin{2 \pi \bm{\phi}} \sin{i} + \lambda_{\rm c},
    \label{eq:planet_atmosphere_wavelength}
\end{equation}
where $i$ is the orbital inclination, $\Delta\lambda_{\rm max}$ the maximum shift in wavelength, $\lambda_{\rm c}$ the absorption wavelength at the centre of the transit, and $\bm{\phi}$ the orbital phase. We modelled the two-dimensional absorption feature of the planetary atmosphere during transit, ${\rm \mathbf{P}(\bm{\phi}, \bm{\lambda}_{\rm obs})}$, as a Gaussian defined as a function of the orbital phase, $\bm{\phi}$, and wavelength, $\bm{\lambda}_{\rm obs}$, through
\begin{equation}
    {\rm \mathbf{P}(\bm{\phi}, \bm{\lambda}_{\rm obs})} = A \cdot \exp{\left(-\frac{(\bm{\lambda}_{\rm obs} - \bm{\lambda}_{\rm p})^2}{2 \sigma^2}\right)}.
    \label{eq:planet_trace_equation}
\end{equation}
$A$ is the depth of the absorption line and $\sigma$ denotes the line width of the Gaussian in \si{\nm}. 
As the planet covers different regions of the stellar surface during transit, performing differential transmission spectroscopy introduces residual spectral lines. This feature is often termed the Doppler shadow, which is a manifestation of the Rossiter-McLaughlin effect in two dimensions. We modelled the stellar residual lines using \StarRotator\footnote{\url{https://github.com/Hoeijmakers/StarRotator}}, a code that calculates the stellar spectrum, $F_{\ast}$, given a certain projected rotational velocity ($v\sin i_\ast$). It determines the spectrum that is obscured by the planet, $\bm{F}_{\rm obsc}(\bm{\phi})$, as a function of the orbital phase, from which the residual spectrum, $\bm{F}_{\rm res}(\bm{\phi})$, is calculated as follows:
\begin{equation}
    \bm{F}_{\rm res}(\bm{\phi}) = \frac{F_{\ast}-\bm{F}_{\rm obsc}(\bm{\phi})}{F_{\ast}}.
    \label{eq:res}
\end{equation}
By default, \StarRotator uses a stellar spectrum from the PHOENIX database \citep{husser_new_2013}, when provided with the effective temperature, $T_{\rm eff}$, the surface gravity, $\log g_\ast$, and the metallicity, $\left[\ch{Fe}/\ch{H}\right]$, of the host star. For this study, we instead modelled the stellar spectrum with \texttt{pySME} \citep{wehrhahn_pysme_2023} based on the VALD line list \citep{piskunov_vald_1995, ryabchikova_major_2015}. For the stellar parameters, we adopted the best-fit values from \citet{prinoth_titanium_2022}, consistent with \citet{lendl_hot_2020} and \citet{deline_atmosphere_2022}. 

\begin{figure}
    \centering
    \includegraphics[width=\linewidth]{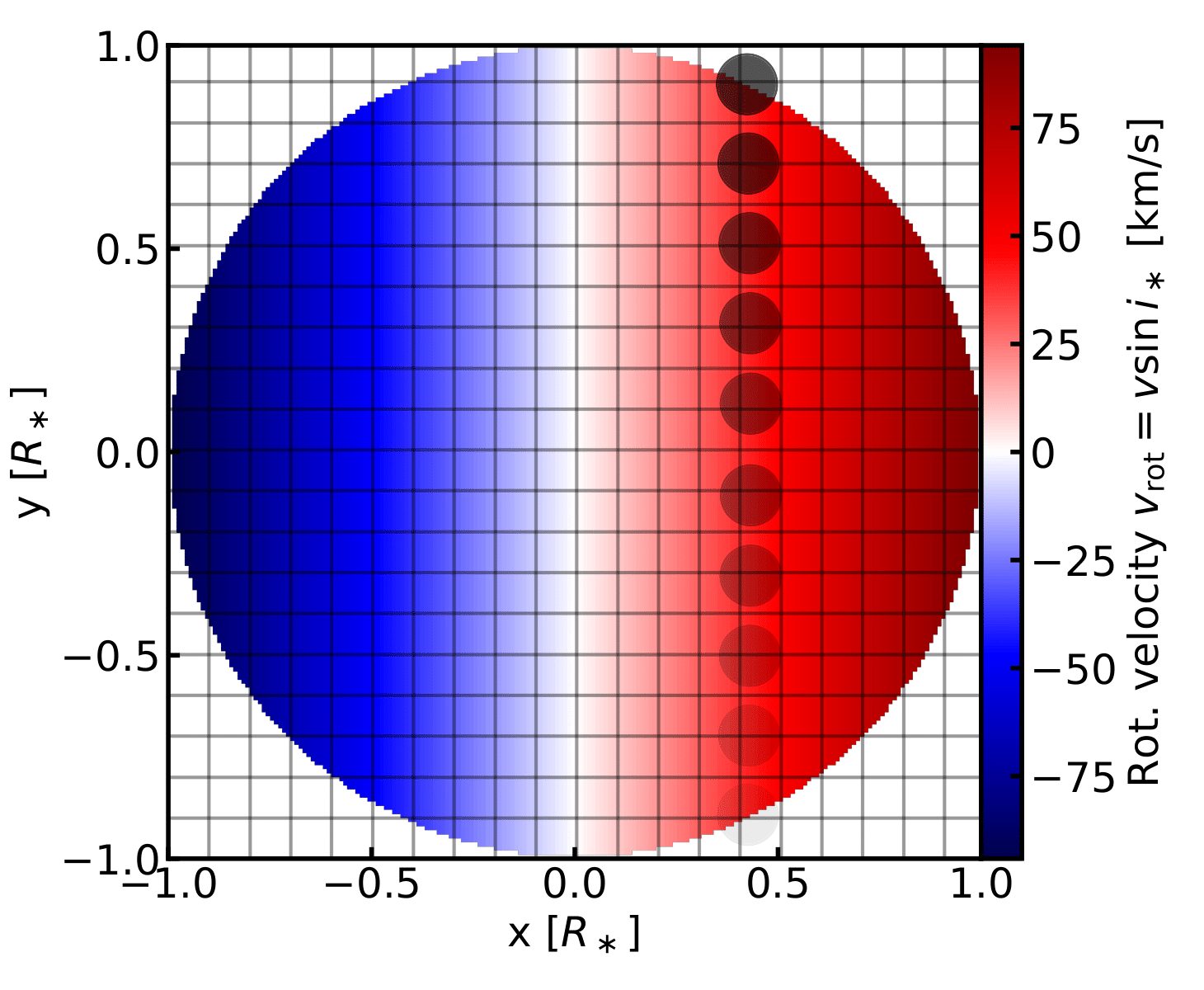}
    \caption{Orbital configuration based on best-fit parameters in Table\,\ref{tab:bestfit}. We used a grid size of 200 for \StarRotator, which means that the whole plane is divided into \num{40000} grid cells (200 in the x direction, 200 in the y direction). Every tenth grid line is indicated in grey. We plot only every fourth position of the planet during the observation sequence, in which the opacity of the planet increases with time to indicate the orbital direction. The rotational velocity, $v_{\rm rot}$, in this case corresponds to $v\sin{i_\ast}$, as we do not fit for the stellar inclination, $i_\ast$. The configuration shows only the positive $\lambda$ scenario.}
    \label{fig:orbital_config}
\end{figure}

To compute the components of the stellar spectrum, \StarRotator divided the stellar surface into a 200\,$\times$\,200 grid of cells of different rotational velocities (see Fig.\,\ref{fig:orbital_config}). We accounted for limb darkening in the stellar spectrum using the quadratic limb-darkening law \citep{kopal_detailed_1950} based on the parameters in \citet{deline_atmosphere_2022}. We assumed no differential rotation, such that grid cells in the vertical direction have the same rotational velocity. The rotational velocity, $\bm{v}_{\rm rot}$, of each cell Doppler-shifts the wavelengths of the observed spectrum, as is described in Eq. \eqref{eq:wl_shift}.

The position of the planet at each observed orbital phase, $\phi$, was calculated as in \citet{cegla_rossiter-mclaughlin_2016}:
\begin{align}
    \bm{x}_{\rm planet} &= \frac{a}{R_\ast} \sin2 \pi \bm{\phi} \\
    \bm{y}_{\rm planet} &= -\frac{a}{R_\ast} \cos2 \pi \bm{\phi} \cos i \\
    \bm{z}_{\rm planet} &= \frac{a}{R_\ast} \cos2 \pi \bm{\phi} \sin i,
\end{align}
where $\frac{a}{R_\ast}$ is the scaled semi-major axis of the system. Because \StarRotator assumed the sky-projected stellar rotation axis to be the y axis, the planet's coordinates needed to be transformed to the same system. This was achieved by using the projected spin-orbit angle, $\lambda$, to rotate the coordinate system:
\begin{align}
    \bm{x}_{\ast, \rm planet} &= \bm{x}_{\rm planet} \cos \lambda - \bm{y}_{\rm planet} \sin \lambda \\
    \bm{y}_{\ast, \rm planet} &= \bm{x}_{\rm planet} \sin \lambda+ \bm{y}_{\rm planet} \cos \lambda \\
    \bm{z}_{\ast, \rm planet} &= \bm{z}_{\rm planet}
\end{align}

\noindent Knowing the planet's position, we calculated the obscured stellar spectrum for Eq.\,\eqref{eq:res}, which was normalised by the median residual for each phase, accounting for flux variations caused by the light curve. \StarRotator and all its additional functionalities will be described in detail in \hoeijmakers and \lam, including the case of elliptical orbits and coupling to \texttt{pySME}. The model, $\bm{M}$, of the planetary trace, $\bm{P}$, and the stellar residual, $ \bm{F}_{\rm res, norm}$, is then given by
\begin{equation}
    \bm{M} = \bm{P} + \bm{F}_{\rm res, norm}.
    \label{eq:model}
\end{equation}

\noindent This model enables the Rossiter-McLaughlin effect and the planetary trace to be fitted simultaneously for the time-resolved absorption line of \ch{Ca+} at $\SI{850}{\nm}$. As is shown in Fig.\,\ref{fig:Ca_plus}, it is not blended with a strong \ch{Fe} line, nor is it a blend between multiple \ch{Ca+} lines of different isotopes, as is the case for the central line at $\SI{854}{\nm}$ \citep[see][for strong blends of isotopes]{kitzmann_mantis_2023}. We note that this blend is indistinguishable due to resolution blurring. Thus, the line at $\SI{850}{\nm}$ is the best candidate to perform the Rossiter-McLaughlin fit, despite being shallower than the other two.

The model, $\bm{M}$, in Eq.\,\eqref{eq:model} uses four fixed parameters: the well-constrained orbital inclination, $i = \num{84.03+-0.14} \deg$ \citep{lendl_hot_2020}; the scaled semi-major axis, $a/R_\ast = $ \numpm{4.600}{+0.031}{-0.025} \citep{lendl_hot_2020}, because of its degeneracy with $R_{\rm p}/R_\ast$; and the limb-darkening parameters, $u_1 = \num{0.41\pm 0.02}$ and $u_2 = \num{0.16\pm 0.03}$, because they are well constrained via the light-curve analysis in \citet{deline_atmosphere_2022}. In our analysis, the limb-darkening parameters are less constrained due to continuum normalisation. The free parameters and priors for our model are given in Table\,\ref{tab:priors}. We sampled from these prior distributions and evaluated the likelihood in a Bayesian framework using a No-U-turn Sampler \citep[see][for a review]{betancourt_convergence_2017}. We implemented this model in \texttt{JAX} and drew posterior samples with \texttt{NumPyro} \citep{jax2018github,bingham_pyro_2018,phan_composable_2019}. We chose 500 warm-up samples and 800 samples over 30 chains, running in parallel. After the completion of the chains, the posterior distributions were analysed and displayed using \texttt{ArviZ} \citep{kumar_arviz_2019} and \texttt{corner} \citep{foreman-mackey_cornerpy_2016}. The best-fit orbital configuration in the coordinate system of \StarRotator is shown in Fig.\,\ref{fig:orbital_config}.

While determining the flux behind the planet over the course of the transit for the Rossiter-McLaughlin effect is not conceptually novel \citep[see e.g.][]{cegla_rossiter-mclaughlin_2016,bourrier_rossitermclaughlin_2021,sicilia_characterization_2022}, the time-resolved signal of \ch{Ca+} enables a fitting procedure in a new setting. Our approach fits and corrects the Rossiter-McLaughlin effect on the stellar spectra instead of doing so in the cross-correlation space via a Gaussian parametrisation. To increase the signal-to-noise, we interpolated both transit time series onto the same phase grid and averaged their contributions for the purpose of fitting the Rossiter-McLaughlin model. This allows access to further parameters that can be determined in our Bayesian framework; in particular, the planet-to-star radius ratio, $R_{p}/R_\ast$, the projected rotational velocity, $v\sin{i_\ast}$, and the projected spin-orbit angle, $\lambda$. In general, our approach can be used for any stellar residual feature strong enough to be seen in the transmission spectrum, for example strong \ch{Fe} lines. 

\begin{figure}
    \centering
    \includegraphics[width=\linewidth]{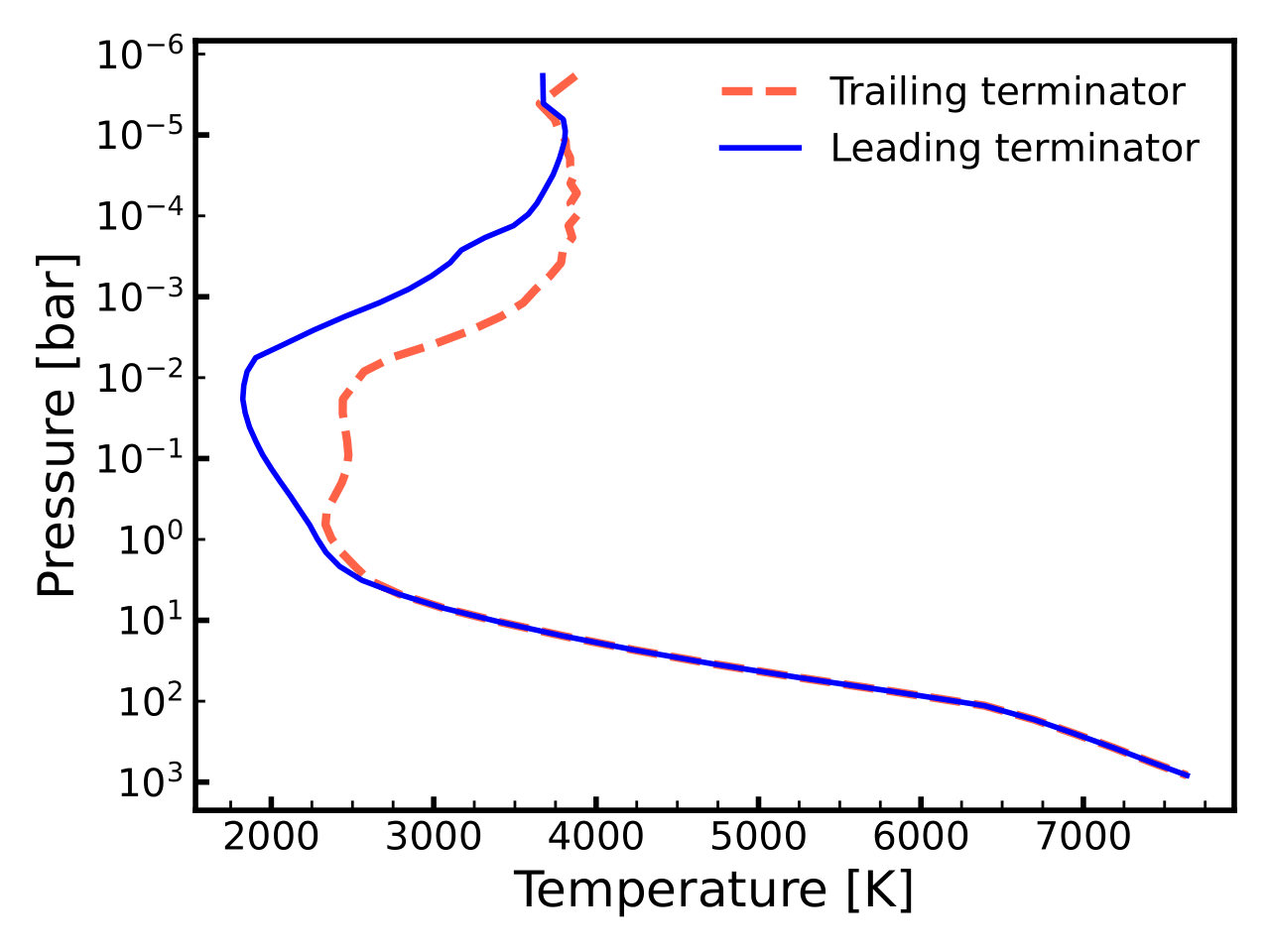}
    \caption{Temperature-pressure profiles for the leading (west) and trailing (east) terminator based on \citet{lee_mantis_2022}.}
    \label{fig:Lee-T-p}
\end{figure}

We corrected the Rossiter-McLaughlin effect over the entire wavelength range using the model in Eq.\,\ref{eq:model} together with the best-fit parameters that describe the stellar component in Table\,\ref{tab:priors}. To remove any systematics in the stellar rest frame before stacking, we vertically de-trended by dividing the mean of each wavelength bin (see \citet{prinoth_titanium_2022, prinoth_time-resolved_2023}). During fitting, we masked the region of the planetary absorption in order to avoid manipulating its signal. To move the spectra to the planetary rest frame, we corrected for the stellar reflex motion caused by the orbiting planet, $v_{\ast, \rm RV}$, and the planetary motion, $v_{\rm p, RV}$, itself by
\begin{equation}
    \bm{v}_{\rm corr} = \bm{v}_{\ast, \rm RV} - \bm{v}_{\rm p, RV},
\end{equation}
where $\bm{v}_{\rm p, RV} = K_{\rm p} \sin{2 \pi \phi}$. We assumed that $K_{\rm p} = v_{\rm orb}\sin{i} = \SI{201}{\km\per\second}$, as was derived in \citet{prinoth_time-resolved_2023}.
\noindent Once in the planetary rest frame, the in-transit exposures were averaged over time, which boosts the signal-to-noise ratio and reveals the planetary absorption feature. Both transit time series were then averaged to increase the signal-to-noise again. To determine the line depth, centre, and width, we fitted a Gaussian function to the observed lines. We further computed atmospheric models for the leading and trailing terminator for comparison, using the temperature-pressure profiles from \citet{lee_mantis_2022}. For both terminators, we averaged the profiles over the observable longitudes between 75.94 and 104.06\,deg (trailing) and between 255.94 and 284.06\,deg (leading). Fig.\,\ref{fig:Lee-T-p} shows the temperature-pressure profiles for both terminators, consistent with the GCM results that the leading terminator is generally colder \citep{lee_mantis_2022}. These models were calculated assuming an equilibrium temperature of \SI{2641\pm 34}{\kelvin} \citep{anderson_wasp-189b_2018}.

For our atmospheric models, we adopted the same procedure as \citet{prinoth_titanium_2022}, with the exception of the isothermal profiles. The planet's atmosphere was assumed to be in chemical and hydrostatic equilibrium, and of solar metallicity. The planetary surface gravity ($g_{\rm p} = \SI{18.8}{\m\per\second\squared}$) and radius ($R_{\rm p} = 1.619\,R_{\rm Jup}$) were adopted from \citet{lendl_hot_2020}, assuming a reference pressure of 10\,bar at the given planetary radius. 
Using the two temperature-pressure profiles in Fig.\,\ref{fig:Lee-T-p}, we then computed the abundance profiles using \texttt{FastChem} \citep{stock_fastchem_2018,stock_fastchem_2022, kitzmann_fastchem_2023}, and further followed the radiative transfer procedure performed in \citet{gaidos_exoplanet_2017}. We used the same opacity functions as \citet{prinoth_titanium_2022}, which include 128 neutral atoms and ions, as well as \ch{H2O}, \ch{TiO}, and \ch{CO}. These opacity functions had previously been computed using \texttt{HELIOS-K} \citep{grimm_helios-k_2015,grimm_helios-k_2021} from the line lists provided by VALD and ExoMol \citep{tennyson_exomol_2016,tennyson_2020_2020,mckemmish_exomol_2019,chubb_exomolop_2021} for atoms and molecules, respectively. 

\section{Results and discussion}
\label{sec:results_discussion}

\subsection{Time-resolved Rossiter-McLaughlin effect}

\begin{figure*}[h!]
    \centering
    \includegraphics[width=\linewidth]{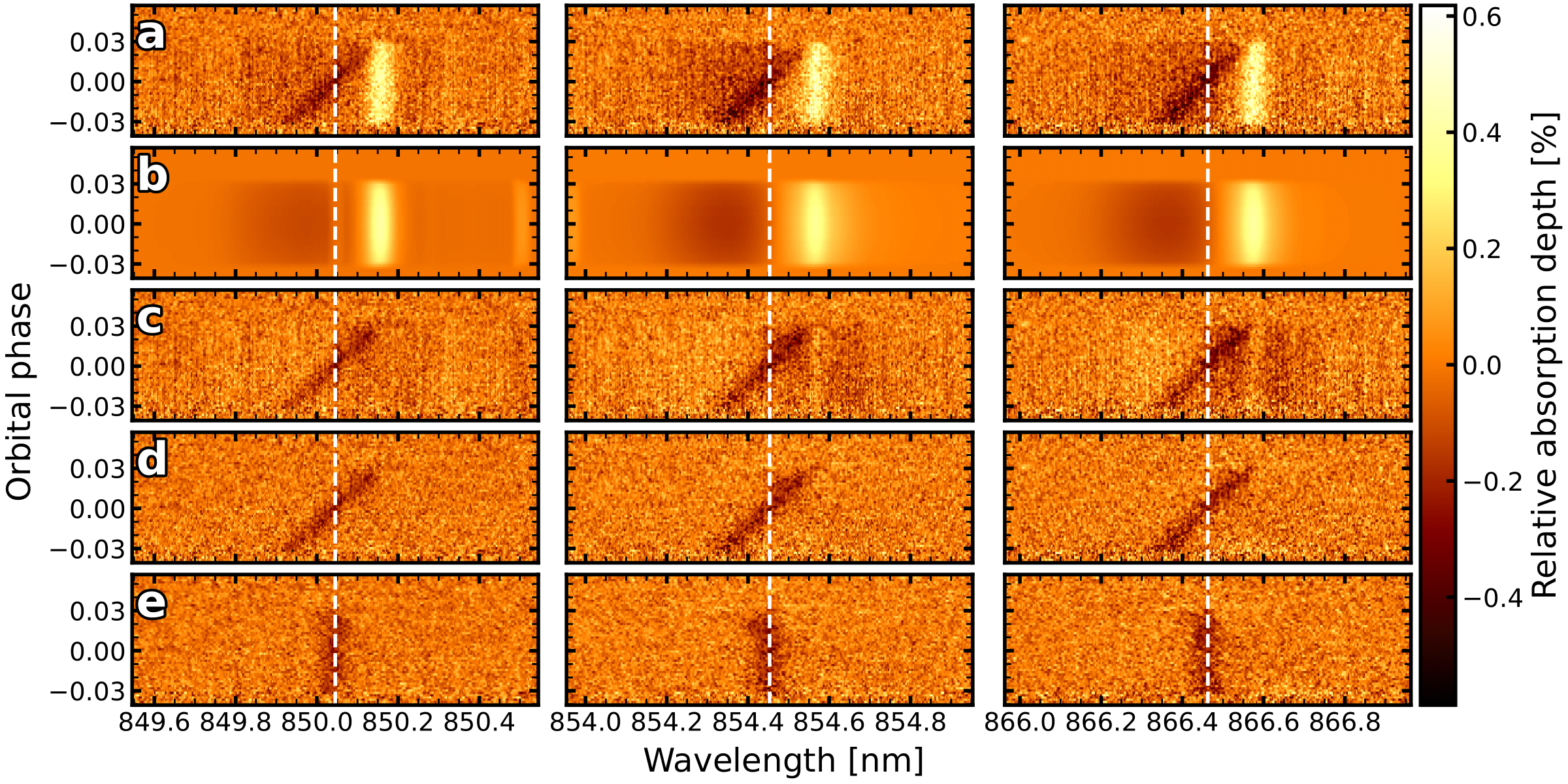}
    \caption{Analysis steps shown for the \ch{Ca+} infrared triplet. \textbf{a:} Combined transmission spectra of both time series in the rest frame of the star. The dark absorption feature is the planetary atmosphere. The bright emission-like feature is the residual originating from the spectra behind the planet during transit (Rossiter-McLaughlin effect). \textbf{b:} Best fit for the stellar residual using the parameters in Table \ref{tab:bestfit}. \textbf{c:} Combined transmission spectrum after dividing out the best fit for the stellar residual. \textbf{d:} Panel c after correcting for systematic noise in the rest frame of the star (vertical de-trending). \textbf{e:} Same as panel d, but in the rest frame of the planet. The vertical absorption features are the \ch{Ca+} lines of the planetary atmosphere.}
    \label{fig:stepbystep}
\end{figure*}


The posteriors of both the planetary and stellar parameters from the fitting of the Rossiter-McLaughlin are shown in Fig.\,\ref{fig:corner} and the resulting median values, including 1$\sigma$ uncertainties, are provided in Table\,\ref{tab:priors}. Fig.\,\ref{fig:stepbystep} shows the best-fit model in the case of the \ch{Ca+} line in Panel b. We used the stellar parameters from Table\,\ref{tab:priors} for \StarRotator to model the residual of the Rossiter-McLaughlin effect for the whole wavelength range, as is described in Section\,\ref{sec:methods}.  Our best-fit parameters for the planet-to-star radius ratio, $R_{\rm p}/R_{\ast}$, the projected spin-orbit angle, $\lambda$, and the projected rotational velocity, $v\sin{i_\ast}$, are largely in agreement with previous studies \citep{anderson_wasp-189b_2018,lendl_hot_2020,deline_atmosphere_2022}. Because our fit includes $v\sin{i_\ast}$, our model cannot distinguish between the scenarios suggested in \citet{deline_atmosphere_2022}, namely $\lambda = 91.7$\,deg and $\lambda = -91.7$\,deg are equally likely, and remain degenerate. 
To this effect, we placed strict uniform priors on $\lambda$ around the positive solution, henceforth ignoring the negative solution and thereby breaking the degeneracy.
\begin{table*}[h!]
    {\centering
    \caption{Planetary and stellar parameters to model the residual of the Rossiter-McLaughlin effect.}
    \label{tab:priors}
    \begin{tabular}{llll}
         \toprule
         \multicolumn{4}{l}{Fitted parameters}  \\ 
         \midrule
          & Symbol [units] & Value & Prior \\ 
         \midrule
         Amplitude of planetary absorption & $A$ [\%] & \num{-1.641+-0.063} & $ \mathcal{U}(-2, -1)$ \\
         Centre wavelength & $\lambda_{\rm p}$ [nm] & \num{850.042+-0.001} & $ \mathcal{U}(850.03 , 850.05)$ \\
         Slope of absorption & $\Delta\lambda$ & \num{0.523+-0.012} & $ \mathcal{U}(0.475, 0.575)$ \\
         Gaussian width of absorption & $\sigma$ [nm] & \num{0.029+-0.001} & $ \mathcal{U}(0.02, 0.04)$ \\
         Planet-to-star radii ratio & $R_{\rm p}/R_{\ast}$ & \num{0.074+-0.001} & $ \mathcal{U}(0.069,0.076)$  \\
         Projected spin-orbit angle & $\lambda$ [$\deg$] & \num{90.07+-0.24} & $ \mathcal{U}(89, 92)$ \\
         Projected rotational velocity & $v\sin{i_\ast}$ [\si{\km\per\second}] & \num{95.05+-0.55} & $ \mathcal{U}(90, 105)$ \\
         \midrule
         \multicolumn{4}{l}{Fixed parameters}  \\ 
         \midrule
          & Symbol [units] & Value & Reference \\ 
          \midrule
         Orbital inclination & $i$ [deg] & \numpm{84.58}{+0.23}{-0.22} & \cite{deline_atmosphere_2022}\\
         Scaled semi-major axis     & $a/R_\ast$ & \numpm{4.600}{+0.031}{-0.025} & \cite{deline_atmosphere_2022}\\
         Limb-darkening parameters & $u_1$ & \numpm{0.414}{+0.024}{-0.022} & \cite{deline_atmosphere_2022} \\
                                   & $u_2$ & \numpm{0.155}{+0.032}{-0.034} & \cite{deline_atmosphere_2022} \\
         \bottomrule
    \end{tabular}\\}
    \textit{Note:} Fitted and fixed parameters for the model in Eq.\,\eqref{eq:model}. All parameters except $A$, $\lambda_{\rm p, vac}$, $\Delta\lambda$, and $\sigma$ were used to model the residual of the Rossiter-McLaughlin effect. 
\end{table*}

\subsection{Spectral atlas of \bibi}
\label{sec:spectral_atlas}

\begin{figure*}[h!]
    \centering
    \includegraphics[width=\linewidth]{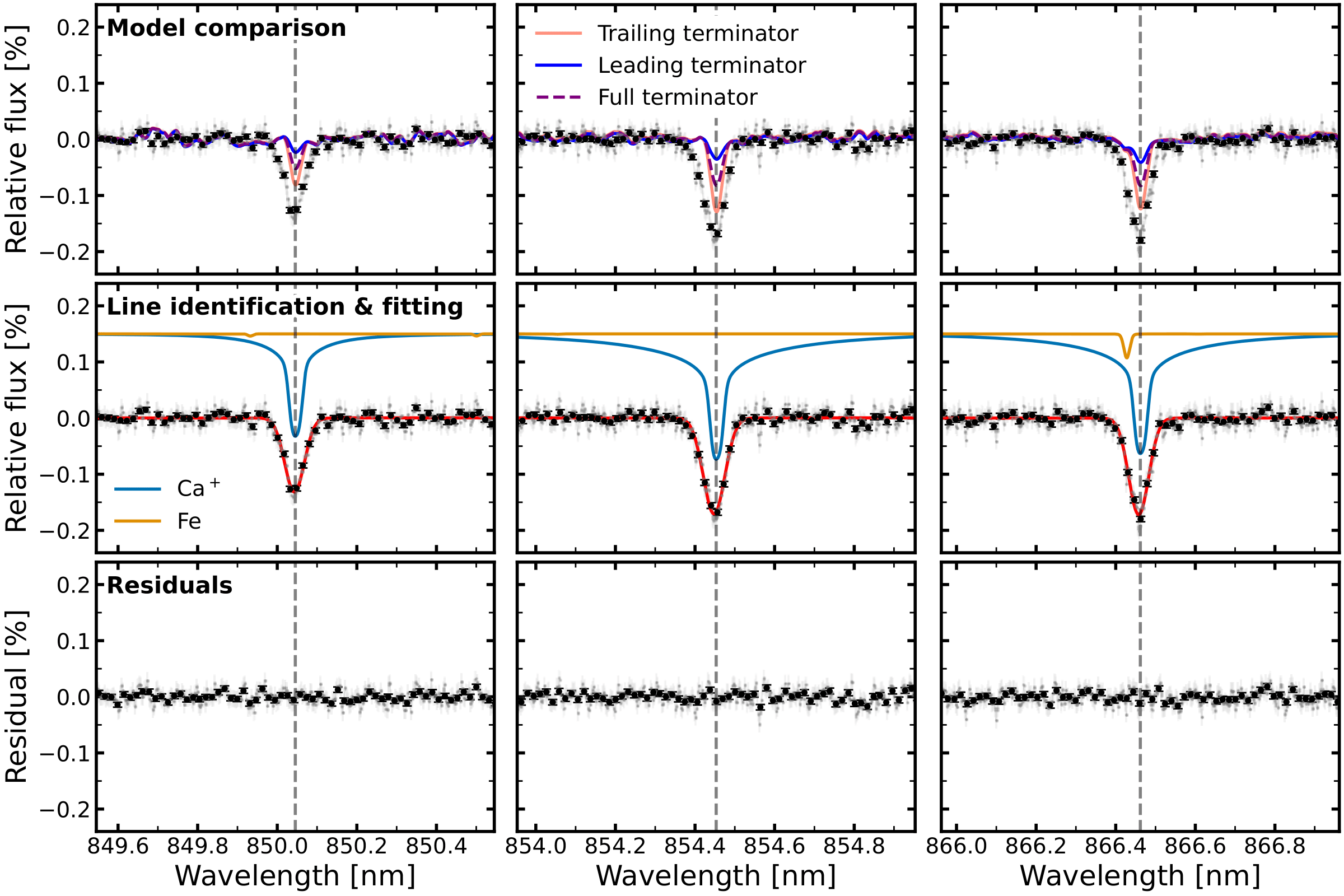}
    \caption{MAROON-X transmission spectrum of \bibi for the \ch{Ca+} infrared triplet stacked in the planetary rest frame for both nights combined. The rest frame transition wavelengths are marked with dashed vertical lines in grey.  The data is shown in grey. The binned data (x8) is shown in black. \textit{Upper panels:} Models for the leading (blue) and trailing (red) terminator computed using the T-p profiles in Fig.\,\ref{fig:Lee-T-p} assuming chemical equilibrium and solar metallicity. The models are sampled at the resolution of the spectrograph ($R \sim \num{85000}$, $v = \SI{3.52}{\km\per\second}$), and additionally broadened to match the planetary rotation \citep[$v_{\rm rot}\sin{i} = \SI{3.04}{\km\per\second}$,][]{prinoth_time-resolved_2023} and exposure smearing ($v_{\rm smear, red} \approx \SI{0.14}{\km\per\second}$). Additionally, the models were continuum-normalised by computing their continuum separately and subtracting it. \textit{Middle panels:} The data has been corrected for tellurics and the residuals of the Rossiter-McLaughlin effect using the parameters in Table\,\ref{tab:bestfit}. The best-fit Gaussians are shown in red. The templates for \ch{Ca+} at \SI{4000}{\kelvin} and \ch{Fe} at \SI{3000}{\kelvin} are shown to compare the line position. \textit{Bottom panels:} Residuals after removing the best fit.}
    \label{fig:Ca_plus}
\end{figure*}

\begin{figure*}
    \centering
    \includegraphics[width=\linewidth]{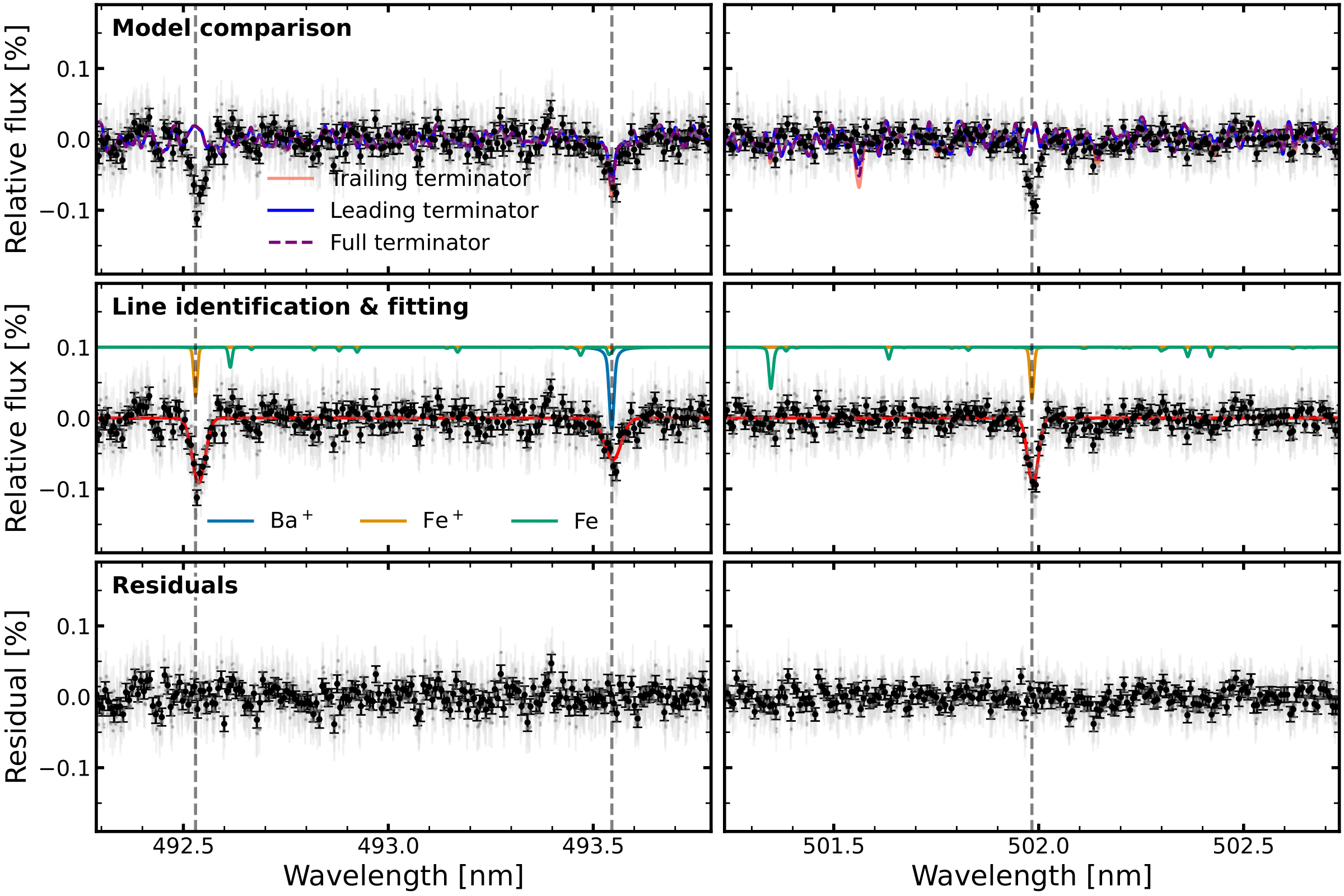}
    \caption{Same as Fig.\,\ref{fig:Ca_plus} but for \ch{Ba+}, \ch{Fe}, and \ch{Fe+}. The templates for atoms and ions are shown at a temperature of \SI{3000}{\kelvin} and \SI{4000}{\kelvin}, respectively. To the right of the \ch{Fe}, there is an unidentified absorption line, while to the left, the model predicts absorption by \ch{Ti}.}
    \label{fig:Ba_plus_Fe_plus_Fe_Ti}
\end{figure*}

\begin{figure*}
    \centering
    \includegraphics[width=\linewidth]{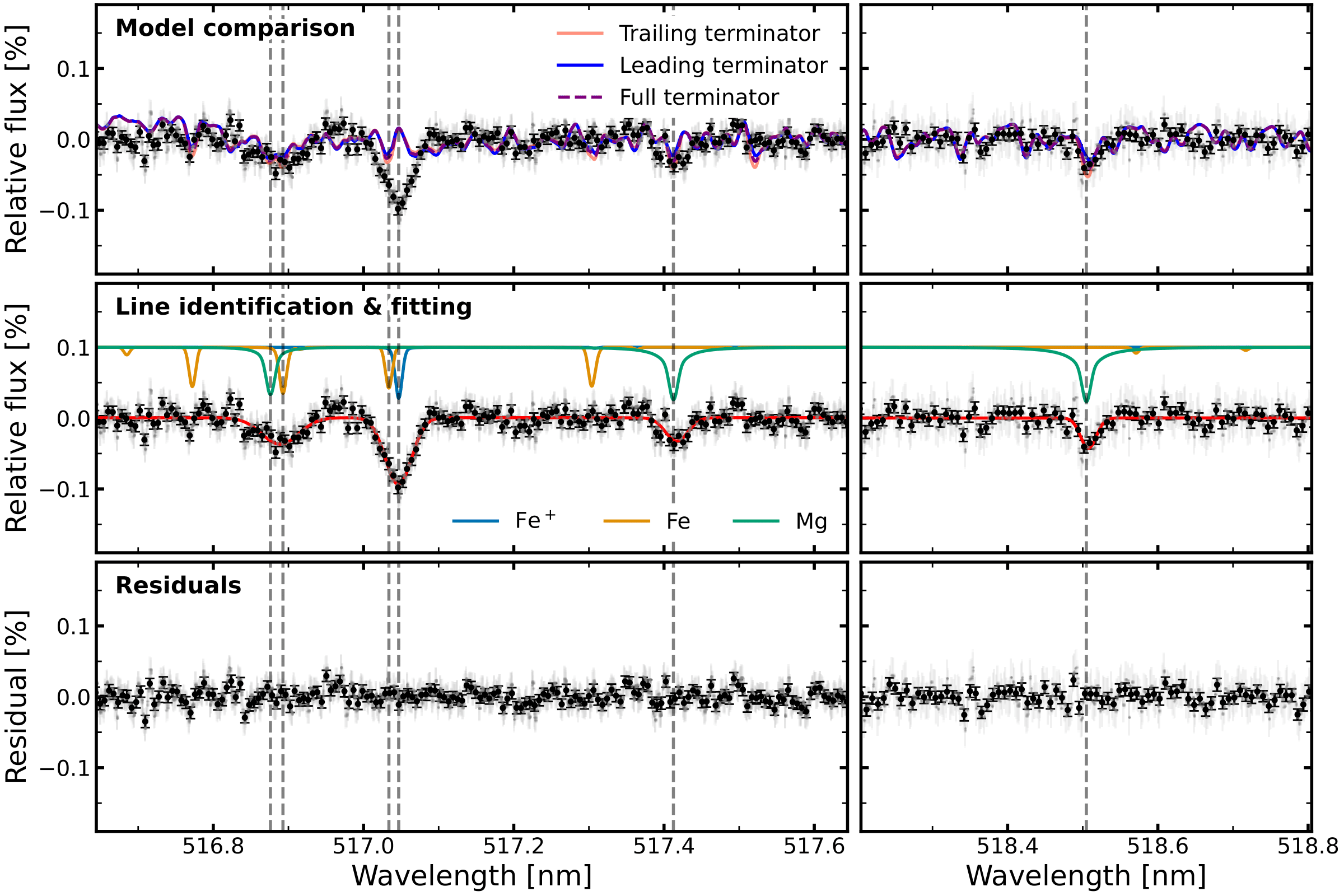}
    \caption{Same as Fig.\,\ref{fig:Ca_plus} but for the \ch{Mg} triplet.}
    \label{fig:Mg}
\end{figure*}

\begin{figure*}
    \centering
    \includegraphics[width=\linewidth]{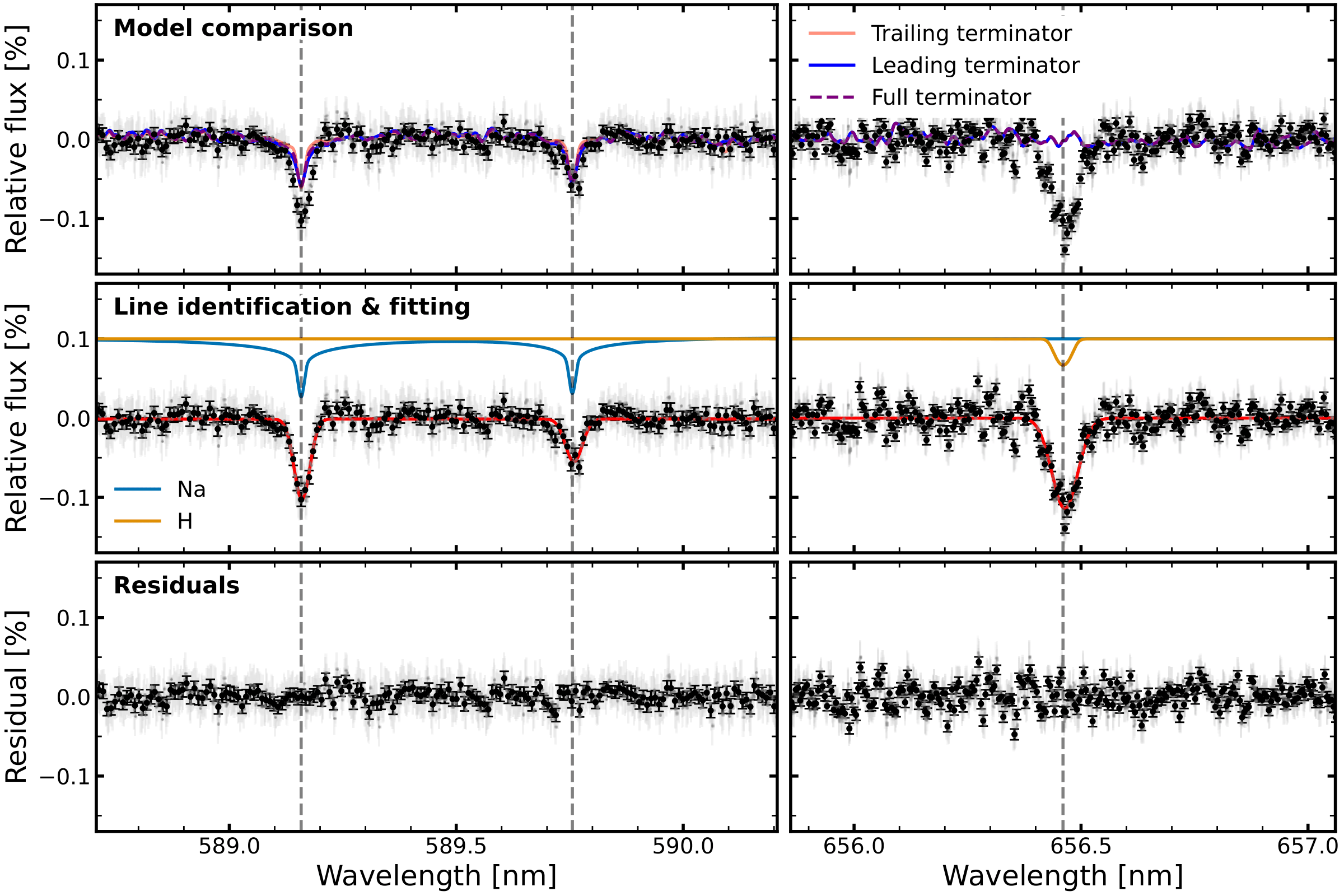}
    \caption{Same as Fig.\,\ref{fig:Ca_plus} but for \ch{Na} and \ch{H$\alpha$}.}
    \label{fig:Na_Halpha}
\end{figure*}

After carrying out the above corrections, we find significant line absorption of various metals in the wavelength range of MAROON-X. The full spectral atlas of \bibi is shown in Appendix\,\ref{app:atlas}. For each of the spectral orders, the final transmission spectrum is shown together with the relevant absorbing species. Additionally, the telluric transmission spectrum is over-plotted to indicate possible regions of heavy telluric contamination. Table\,\ref{tab:bestfit} shows the best-fit Gaussian parameters of a few strong absorbers. In light of the richness of the transmission spectrum of \bibi, we have only selected a set of strong lines to show the potential for this spectrum and discuss implications for its chemistry. 

We detect strong absorption from the \ch{Ca+} infrared triplet at 38.5$\sigma$, 63.1$\sigma$, and 65.9$\sigma$ (see Table\,\ref{tab:bestfit}). As is seen in Fig.\,\ref{fig:Ca_plus}, the forward models for the two terminators do not match the observed line depth, which points towards deviations from local thermal equilibrium (LTE) and shortcomings of current atmospheric models. Future analysis in the form of retrievals is required to determine the temperature-pressure profile.

Apart from \ch{Ca+}, we also detect line absorption from \ch{Ba+}, \ch{Na}, \ch{Mg}, \ch{Fe+}, \ch{Fe}, and \ch{H$\alpha$} at above the 5$\sigma$ detection threshold that we adopted, details of which are provided in Table\,\ref{tab:bestfit} and shown in Figures\,\ref{fig:Ca_plus}--\ref{fig:Na_Halpha}. All the detected species have previously been reported using the cross-correlation technique \citep{stangret_high-resolution_2022,prinoth_titanium_2022,prinoth_time-resolved_2023}, which, together with detections from this analysis, corroborates their presence.
The detection of \ch{Fe+} is surprising given that its presence is not predicted by the models, which could indicate that the transmission spectrum probes regions of the atmosphere that are hotter than \SI{3000}{\kelvin} where \ch{Fe+} absorption becomes significant \citep{prinoth_time-resolved_2023} or that models beyond our current assumptions of local thermodynamic and hydrostatic or chemical equilibrium are required to explain the observed absorption, especially at higher altitudes, through for example photoionisation 
\citep{FisherHeng2019, Brogi2021,zhang_detection_2022}. 

Atomic absorption by \ch{Mg} and \ch{Fe} is predicted by the model, but not as strongly as observed in the data. The \ch{Ba+} absorption at $\SI{493.55}{\nm}$, which is blended with weak \ch{Fe} absorption, indeed seems to be roughly consistent with the observed depth at the hotter, trailing terminator. Only a few lines contribute to the observed cross-correlation signal of \ch{Ba+} in the wavelength range of MAROON-X, where most of them appear to be blended with \ch{Fe} or \ch{Fe+} lines. This line at $\SI{493.55}{\nm}$ may be subject to a partial blend with a weak \ch{Fe+} line (see Fig.\ref{fig:Ba_plus_Fe_plus_Fe_Ti}). Another \ch{Ba+} line with no \ch{Fe} blend is expected at $\SI{585.5}{\nm}$, which is observed, while the one at $\SI{614.}{\nm}$ is overlapping with an \ch{Fe} line, which is also observed. All these lines are shown in Appendix \ref{app:atlas} in Fig.\,\ref{fig:atlas_0_blue} and Fig\,\ref{fig:atlas_0_red}. Nevertheless, other spectral lines may be contributing in unison and significantly affecting the overall line strength from aliasing effects, when using the cross-correlation technique \citep{borsato_mantis_2023}.\\

Fig.\,\ref{fig:Na_Halpha} shows the absorption by the \ch{Na} D-lines and H$\alpha$. \citet{langeveld_survey_2022} surveyed six ultra-hot Jupiters for narrow-band absorption of sodium, among which was also \bibi. In particular, our detection of sodium agrees with the absorption of sodium presented in that previous study, in that the D1 line is shallower than the D2 component. A difference in absorption depths between the D1 and D2 lines could result from a large fraction of the absorbing sodium being optically thin and not following a hydrostatic number density profile \citep{hoeijmakers_high-resolution_2020}. This could be the case if sodium exists in an optically thin torus \citep{oza_sodium_2019, gebek_alkaline_2020} or a tenuous hydrodynamically escaping envelope \citep{wyttenbach_mass-loss_2020}.



\citet{sreejith_cute_2023} observed \bibi with the \textit{CUTE} satellite, detecting \ch{Mg+} lines with absorption depths larger than the Roche Lobe at Lagrange Point 1 (L1) in the near-ultraviolet. Assuming the continuum to be at one planetary radius, we expect the Roche Lobe to be at $R_{\rm p}\approx$ 1.1463, suggesting that the \ch{Ca+} IRT lines probe close to the expected altitude of L1, and thus probe the exosphere.

We note that differences in line positions in comparison to previous works \citep{prinoth_titanium_2022, prinoth_time-resolved_2023} may indeed come from assuming a different value for the orbital velocity, $K_p$. In \citet{prinoth_time-resolved_2023}, $K_p$ was treated as a free parameter for individual species, while this study assumes a value of $K_{\rm p} = \SI{201}{\km/s}$, derived from the orbital period, throughout the entirety of the spectral range, treating all the species equally. This may indeed introduce shifts as different species likely probe different altitudes, and hence different dynamical regimes, which can manifest themselves as different orbital velocities.

\subsubsection{Implications for high-resolution retrievals}

The two MAROON-X transits in this study provide a benchmark data set for retrieval studies at high spectral resolution and are expected to facilitate testing models of atmospheric chemistry and dynamics in these kinds of planets. While the centres of the observed absorption lines seem to be consistent with the rest frame positions when assuming an orbital velocity of $K_p \approx \SI{201}{\km\per\second}$, the lines from distinct species show different broadening, possibly caused by different pressure levels at which different lines or species are probed or by atmospheric dynamics. A full dynamics retrieval is beyond the scope of this paper, but the data set certainly invites further investigation for both dynamical and composition retrievals at high spectral resolution \citep[e.g.][]{brogi_retrieving_2019,gibson_relative_2022,pelletier_vanadium_2023}. 

\begin{table*}[h!]
    \caption{Gaussian fits of the planetary absorption lines sorted by wavelength.}
    {\centering
    \small
    \begin{tabular}{ll|ccccccc}
        \toprule
        &&  $A$ [ppm] & $\lambda_{\rm centre}$ [nm] & $\sigma_{\rm Gauss}$ [nm] &  $v_{\sigma, \rm Gauss}$ [\si{\km\per\second}] & $\Delta v_{\rm centre }$ [\si{\km\per\second}] & $\sigma$ & Eq. atm. height $h$ [$R_{\rm p}$]\\
        \midrule
        \ch{Fe+}        & $492.53$  & $918 \pm 79$  & $492.537 \pm 0.002$ & $0.016 \pm 0.002$ & $9.69 \pm 0.96$     & $4.62 \pm 0.96$   & $11.6$    &  $1.0806 \pm 0.0070 $ \\
        \ch{Ba+}        & $493.55$  & $591 \pm 68$  & $493.548 \pm 0.003$ & $0.020 \pm 0.003$ & $12.24 \pm 1.63$    & $1.55 \pm 1.63$   & $8.7$     &  $1.0526 \pm 0.0061 $ \\
        \ch{Fe+}        & $501.98$  & $929 \pm 72$  & $501.986 \pm 0.001$ & $0.013 \pm 0.001$ & $7.73 \pm 0.69$     & $1.47 \pm 0.69$   & $12.9$    &  $1.0815 \pm 0.0064 $ \\
        \ch{Mg}         & $517.41$  & $331 \pm 25$  & $517.417 \pm 0.001$ & $0.015 \pm 0.001$ & $8.70 \pm 0.77$     & $2.53 \pm 0.76$   & $13.2$    &  $1.0296 \pm 0.0024 $ \\
        \ch{Mg}         & $518.50$  & $427 \pm 76$  & $518.507 \pm 0.002$ & $0.011 \pm 0.002$ & $6.08 \pm 1.25$     & $1.55 \pm 1.25$   & $5.6$     &  $1.0383 \pm 0.0068 $ \\
        \ch{Na}         & $589.16$  & $1016 \pm 39$ & $589.160 \pm 0.001$ & $0.017 \pm 0.001$ & $8.87 \pm 0.39$     & $0.88 \pm 0.4$    & $26.1$    &  $1.0888 \pm 0.0040 $ \\
        \ch{Na}         & $589.76$  & $540 \pm 37$  & $589.759 \pm 0.001$ & $0.019 \pm 0.001$ & $9.49 \pm 0.75$     & $1.74 \pm 0.75$   & $14.6$    &  $1.0482 \pm 0.0035 $ \\
        \ch{Ba+}        & $614.34$  & $346 \pm 27$  & $614.341 \pm 0.003$ & $0.030 \pm 0.003$ & $14.74 \pm 1.35$    & $-0.11 \pm 1.34$  & $12.8$    &  $1.0311 \pm 0.0026 $ \\
        \ch{H$\alpha$}  & $656.46$  & $1184 \pm 23$ & $656.464 \pm 0.001$ & $0.028 \pm 0.001$ & $12.83 \pm 0.3$     & $1.55 \pm 0.30$   & $51.5$    &  $1.1029 \pm 0.0033 $ \\
        \ch{Ca+}        & $850.05$  & $1322 \pm 38$ & $850.043 \pm 0.001$ & $0.025 \pm 0.001$ & $8.68 \pm 0.29$     & $-0.78 \pm 0.29$  & $34.8$    &  $1.1136 \pm 0.0043 $ \\
        \ch{Ca+}        & $854.45$  & $1711 \pm 31$ & $854.448 \pm 0.001$ & $0.027 \pm 0.001$ & $9.51 \pm 0.20$     & $-1.89 \pm 0.20$  & $55.2$    &  $1.1449 \pm 0.0044 $ \\
        \ch{Ca+}        & $866.46$  & $1723 \pm 29$ & $866.458 \pm 0.001$ & $0.025 \pm 0.001$ & $8.70 \pm 0.17$     & $-1.41 \pm 0.17$  & $59.4$    &  $1.1458 \pm 0.0044 $ \\
       \bottomrule
    \end{tabular}\\
    } \vspace{1em}
    \textit{Note:} $A$ is the absorption depth in ppm, $\lambda_{\rm centre}$ is the central wavelength of the absorption in nm, $\sigma$ is the Gaussian width of the absorption in nm, $v_{\sigma, \rm Gauss}$ is the Gaussian width of the absorption in \si{\km\per\second}, $\Delta v_{\rm centre }$ is the offset from the expected centre wavelength in \si{\km\per\second}, $\sigma$ is the detection significance, calculated from the absorption depth, $A$, and its uncertainty, and $h$ is the approximate equivalent atmospheric height in units of $R_{\rm p}$, calculated as $h \approx \frac{R_\ast}{R_{\rm p}} \sqrt{\left(\frac{R_{\rm p}}{R_\ast}\right)^2 + A}$ \citep{pino_diagnosing_2018}. We note that this assumes the continuum to be at 1\,$R_{\rm p}$. The uncertainties in $A$, $\lambda_{\rm centre}$, and $\sigma_{\rm Gauss}$ are provided by \texttt{lmfit}. The uncertainties for all other parameters were calculated assuming Gaussian error propagation.
    \label{tab:bestfit}
\end{table*}

\subsection{High-resolution spectroscopy with the Extremely Large Telescope}
\begin{figure}
    \centering
    \includegraphics[width=\linewidth]{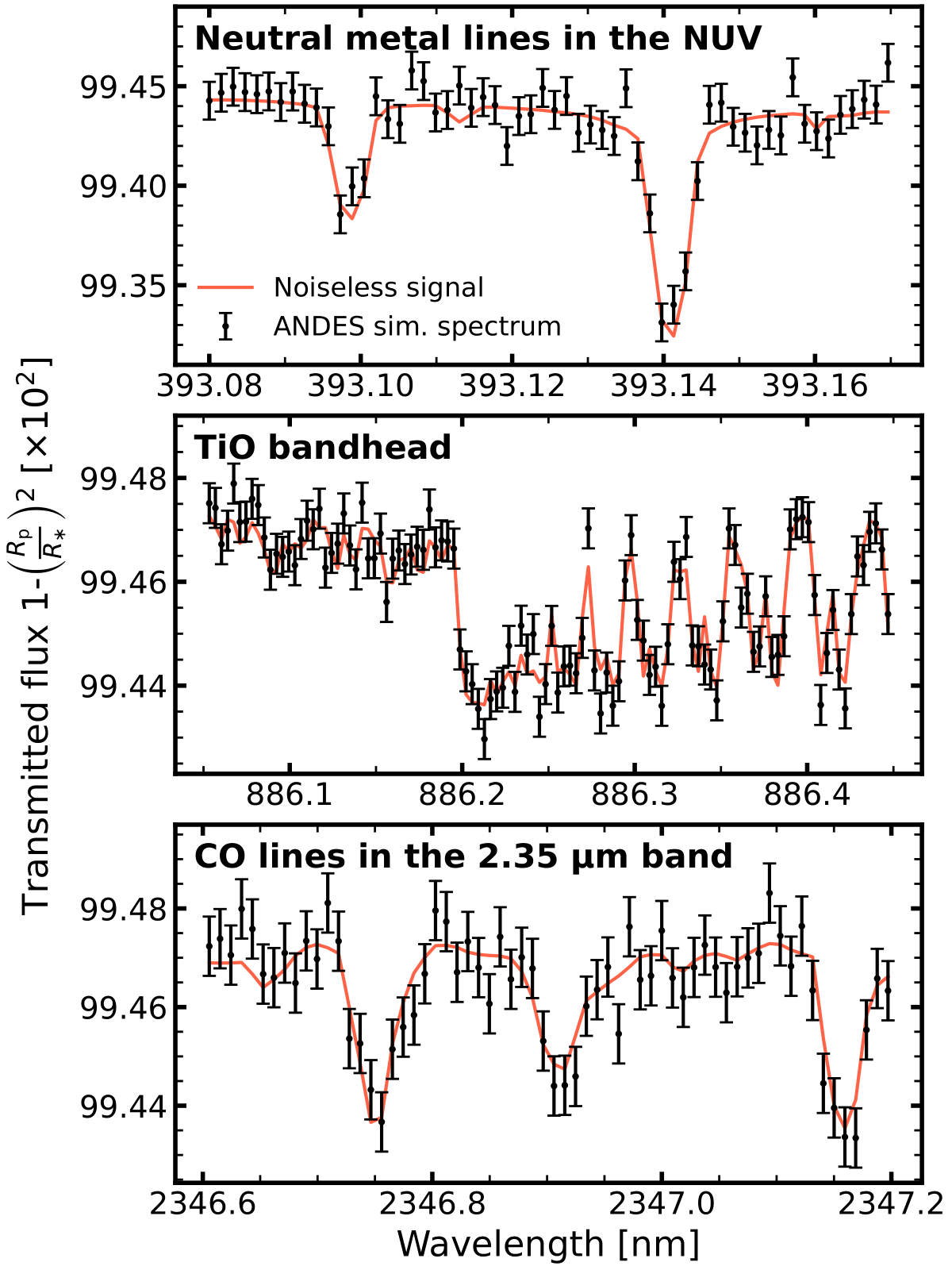}
    \caption{Simulated transmission spectrum of WASP-189\,b, as observed by ANDES based on the 2500\,K equilibrium chemistry model \citep{prinoth_titanium_2022}, using version 1.1 of the ANDES ETC. The three panels show three different wavelength ranges within the ANDES waveband from 0.35 to 2.5 $\mu$m, with single metal lines (top panel), a TiO band-head (middle panel), and CO lines (bottom panel).}
    \label{fig:andes}
\end{figure}

The rich absorption spectra of bright transiting ultra-hot Jupiter systems will be highly amenable to detailed, resolved, single-line spectroscopy with high-resolution spectrographs on the Extremely Large Telescope (ELT), in particular ANDES, covering wavelengths from the near-ultraviolet to the K band. We have used version 1.1 of the ANDES ETC \footnote{\url{http://tirgo.arcetri.inaf.it/nicoletta/etc_andes_sn_com.html}, July 2023}, and the isothermal 2500\,K equilibrium chemistry model \citep{prinoth_titanium_2022} to simulate the entire transmission spectrum of WASP-189\,b after a single transit observed with ANDES, assuming perfect correction of the telluric spectrum and the Rossiter-McLaughlin effect. The ANDES ETC calculates the expected signal-to-noise ratio at a single wavelength, and so we interpolate between the centres of the B, V, J, H, and K bands, where the magnitudes are known \citep{Hog2000,Cutri2003}. We further assume an exposure time of 180\,s and an out-of-transit baseline equal to the transit duration. The resulting spectrum spans from 0.35 to 2.5\,$\mu$m, and examples of well-resolved metal, TiO, and CO lines are shown in Fig.\,\ref{fig:andes}. Based on this simulation, it is expected that ANDES will have the ability to directly resolve a large variety of elements and molecules in the atmospheres of ultra-hot Jupiters. In particular, it will be highly sensitive to the valuable C/O ratio, which is indicative of formation channels \citep{oberg_effects_2011}.

\section{Conclusions}
\label{sec:conclusion}

High-resolution transmission spectroscopy has entered the era of time-resolved signals for single-line spectroscopy \citep[see e.g.][]{pino_neutral_2020,seidel_hot_2022}, cross-correlation analyses \citep[see e.g.][]{kesseli_atomic_2022,prinoth_time-resolved_2023}, and retrieval studies \citep[see e.g.][]{gandhi_retrieval_2023}. In our work, we present a detailed analysis of the transmission spectrum of \bibi observed with MAROON-X, using a time-resolved approach to fit the stellar and planetary components. We fitted the stellar residual imposed by the planet covering parts of the stellar disc during transit, combining \texttt{pySME}, \StarRotator, \texttt{numpyro}, and \texttt{JAX} using a Bayesian framework to infer posterior distributions of the parameters to model the Rossiter-McLaughlin effect. Our best-fit parameters are in agreement with the values found in previous studies of radial velocity data \citep{anderson_wasp-189b_2018} and light-curve analyses \citep{lendl_hot_2020,deline_atmosphere_2022}. 

The high signal-to-noise data observed with MAROON-X allows for a detailed study of the transmission spectrum of \bibi, revealing single-line absorption at high significance of a variety of chemical species previously detected using the cross-correlation technique \citep{stangret_high-resolution_2022, prinoth_titanium_2022, prinoth_time-resolved_2023}, notably \ch{Ca+}, \ch{Na}, \ch{H$\alpha$}, \ch{Mg}, \ch{Fe}, and \ch{Fe+}. This data set provides a significant step forward in characterising exoplanetary atmospheres in unprecedented detail.

These observations of \bibi with MAROON-X provide a benchmark data set for high-resolution retrieval studies for both composition and dynamics. As is seen from our model comparison, deviations from local thermal equilibrium are to be considered in order to explain the observed absorption lines. It is becoming evident, yet again, that one-dimensional models do not suffice in reproducing the observed transmission spectrum and three-dimensional retrievals at high spectral resolution are required to explain the observed line depths to reflect the true nature of these planets.

\begin{acknowledgements}
B.P., H.J.H., and N.W.B acknowledge partial financial support from The Fund of the Walter Gyllenberg Foundation. S.P. acknowledges financial support from the Natural Sciences and Engineering Research Council (NSERC) of Canada and the Fond de Recherche Québécois-Nature et Technologie (FRQNT; Québec). B.T.\ acknowledges the financial support from the Wenner-Gren Foundation (WGF2022-0041).
This work is based on data obtained at the international Gemini Observatory, a programme of NSF’s NOIRLab. The international Gemini Observatory at NOIRLab is managed by the Association of Universities for Research in Astronomy (AURA) under a cooperative agreement with the National Science Foundation on behalf of the Gemini partnership: the National Science Foundation (United States), the National Research Council (Canada), Agencia Nacional de Investigación y Desarrollo (Chile), Ministerio de Ciencia, Tecnología e Innovación (Argentina), Ministério da Ciência, Tecnologia, Inovações e Comunicações (Brazil), and Korea Astronomy and Space Science Institute (Republic of Korea). The MAROON-X team acknowledges funding from the David and Lucile Packard Foundation, the Heising-Simons Foundation, the Gordon and Betty Moore Foundation, the Gemini Observatory, the NSF (award number 2108465), and NASA (grant number 80NSSC22K0117). We thank the staff of the Gemini Observatory for their assistance with the commissioning and operation of the instrument.
This work was enabled by observations made from the Gemini North telescope, located within the Maunakea Science Reserve and adjacent to the summit of Maunakea. We are grateful for the privilege of observing the Universe from a place that is unique in both its astronomical quality and its cultural significance.
\end{acknowledgements}

\newpage

\begin{appendix}
\onecolumn
\newpage
\section{Corner plots}
\begin{figure*}[h!]
    \centering
    \includegraphics[width=\linewidth]{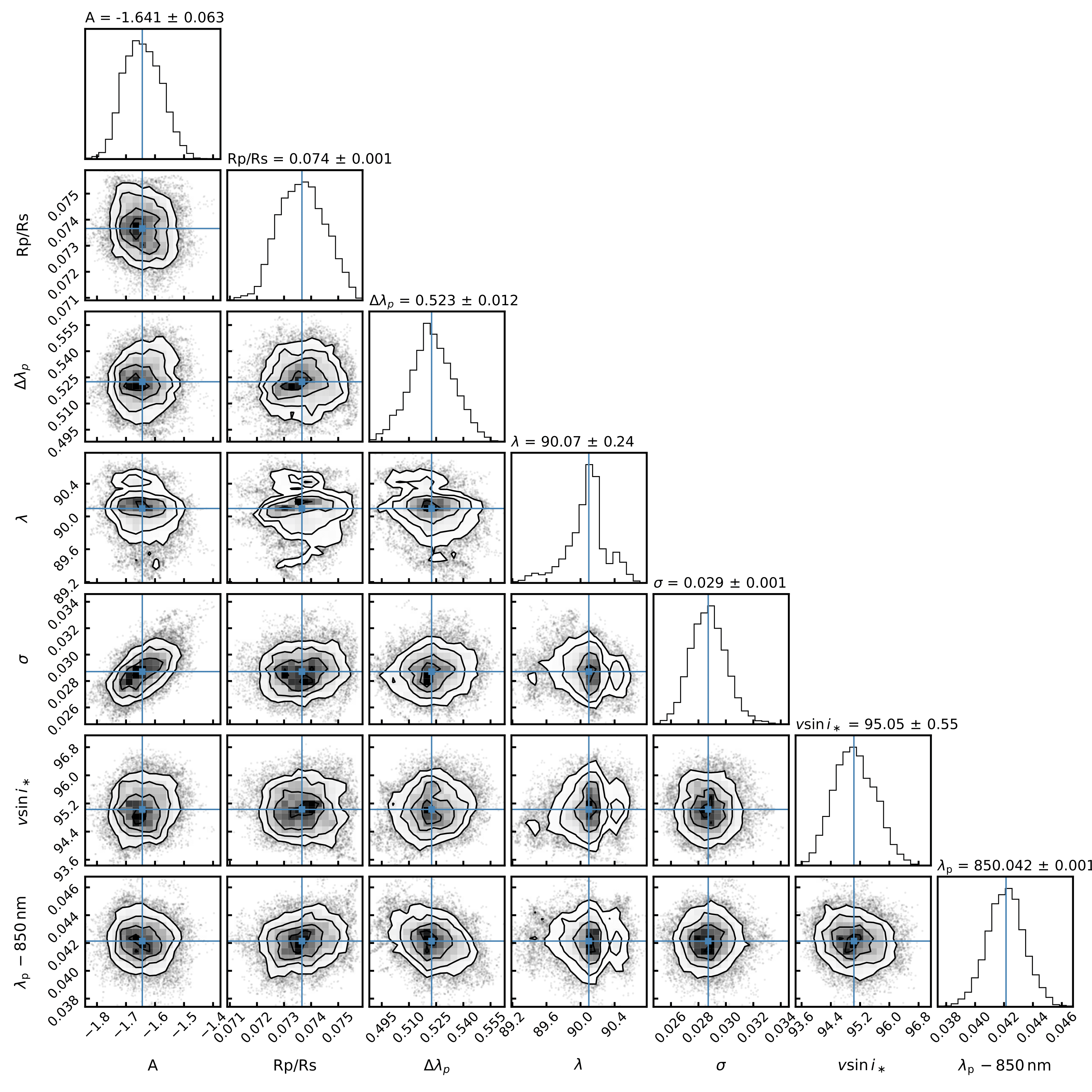}
    \caption{Posterior distribution of the model parameters of Eq. \eqref{eq:model} for the \ch{Ca+} line at $\lambda_{\rm p} = \SI{850.042}{\nm}$. The blue lines indicate the median values of the posterior distribution. The amplitude, $A$, is in units of parts per thousand ($\times 10^{-3}$), the maximum wavelength shift, $\Delta \lambda_{\rm p}$, in nm, the spin-orbit angle, $\lambda$, in deg, the Gaussian width, $\sigma$, in nm, the projected rotational velocity, $v \sin i_\ast$, in \si{\km\per\second}, and the centre wavelength of the absorption, $\lambda_{\rm p}$, in nm. }
    \label{fig:corner}
\end{figure*}



\section{Spectral atlas}
\label{app:atlas}

\begin{sidewaysfigure}
    \centering
    \subfloat{\includegraphics[width=\linewidth]{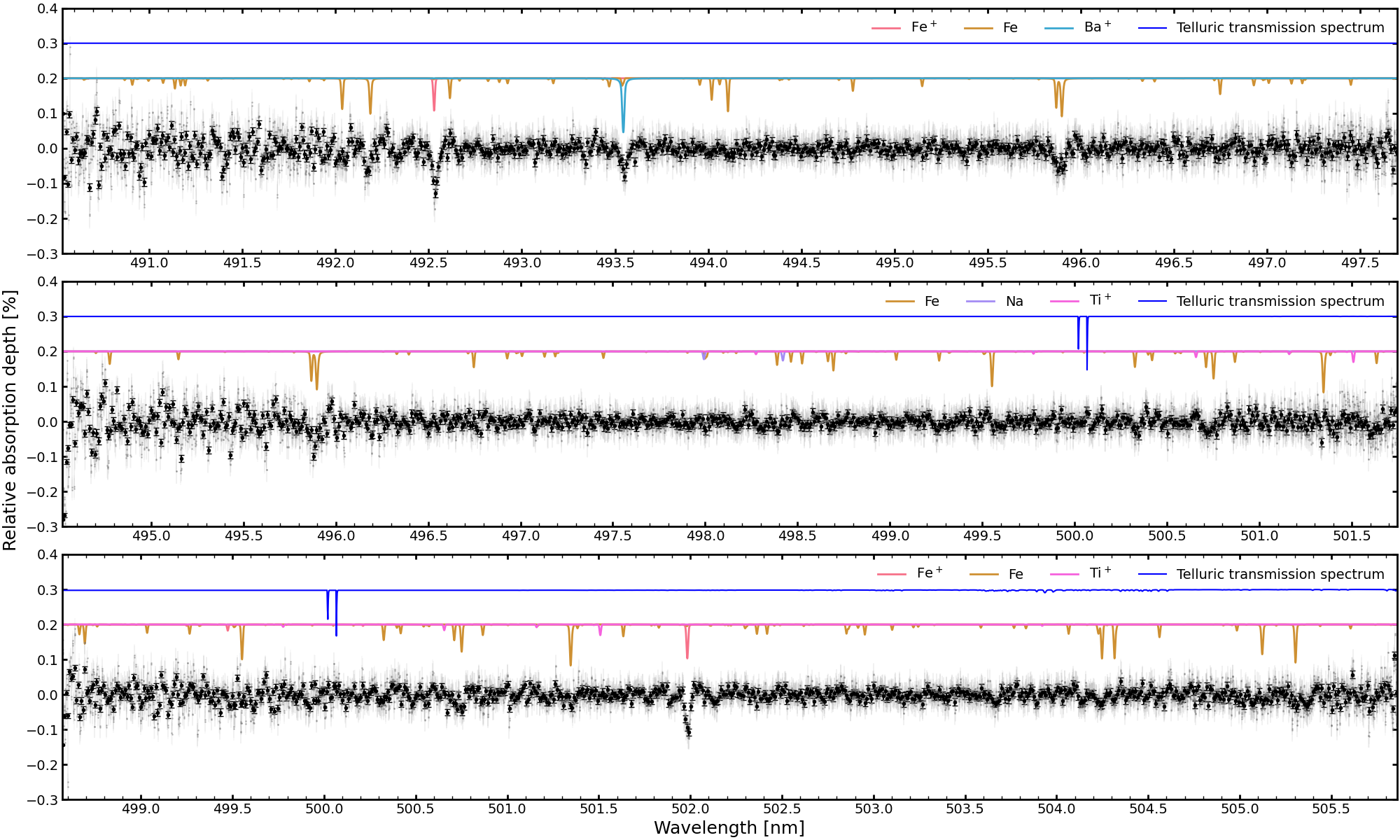}}
    \caption{Spectral atlas of the blue arm of MAROON-X. The grey data points show the unbinned transmission spectrum for each spectral order. The black data points provide the binned spectrum (x10). Additionally, the templates with significant absorption lines are plotted to identify the absorption lines in the transmission spectrum. The templates have been broadened to fit the resolution of the spectrograph. The dark blue spectrum shows a model telluric transmission spectrum to identify heavily contaminated regions.}
    \label{fig:atlas_0_blue}
\end{sidewaysfigure}

\begin{sidewaysfigure}
    \centering
    \ContinuedFloat
    \subfloat{\includegraphics[width=\linewidth]{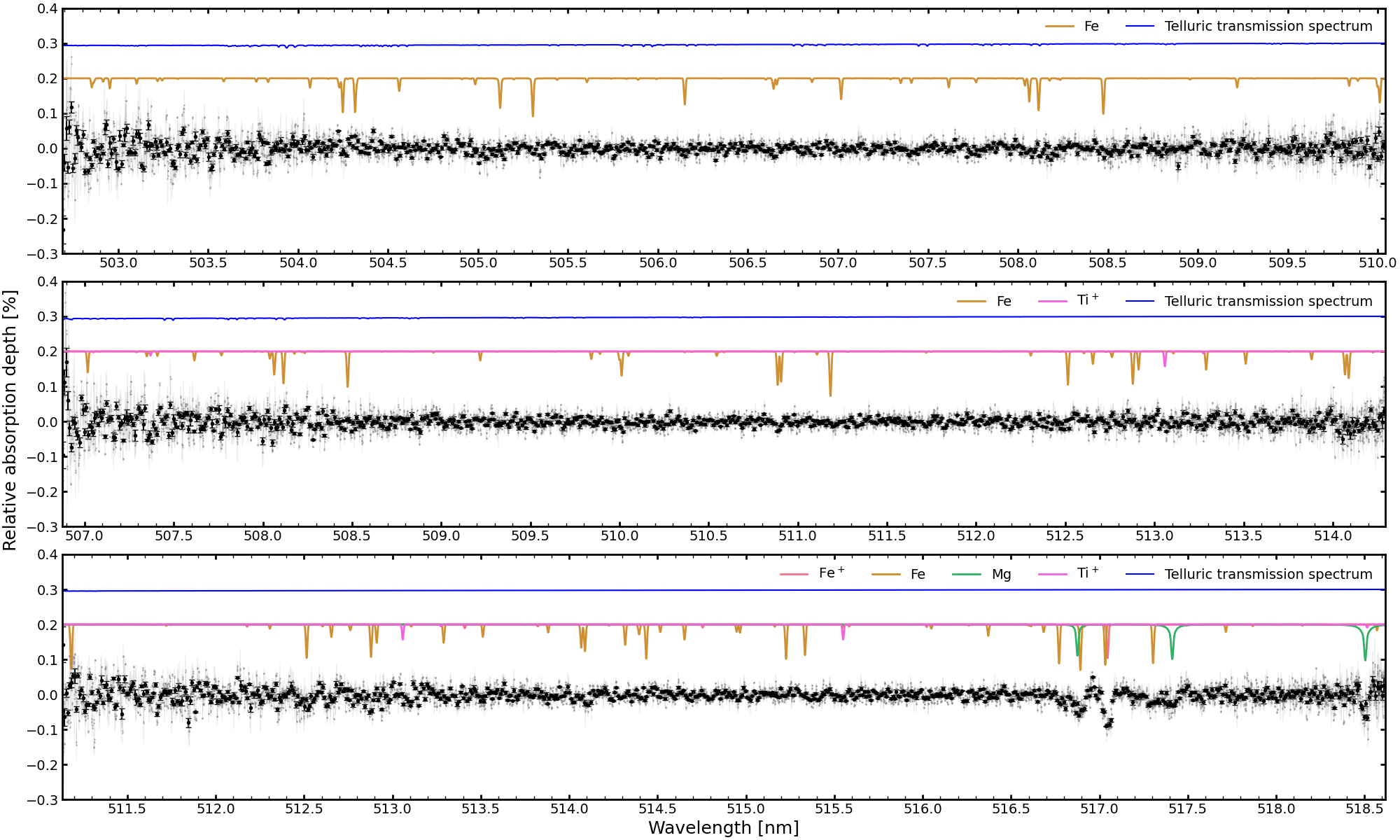}}\\[1em]
    Fig.\,\ref{fig:atlas_0_blue} continued.
\end{sidewaysfigure}

\begin{sidewaysfigure}
    \centering
    \ContinuedFloat
    \subfloat{\includegraphics[width=\linewidth]{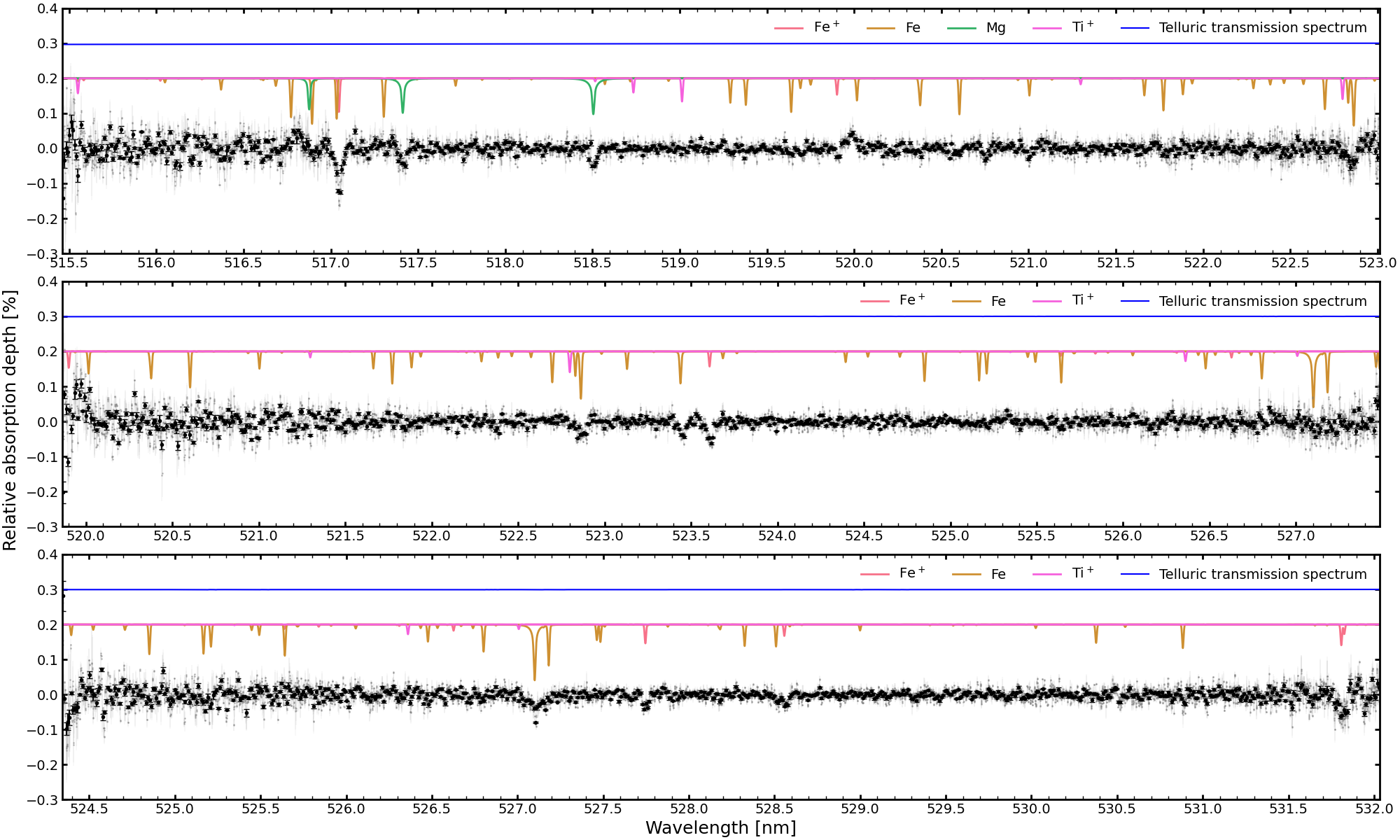}}\\[1em]
    Fig.\,\ref{fig:atlas_0_blue} continued.
\end{sidewaysfigure}

\begin{sidewaysfigure}
    \centering
    \ContinuedFloat
    \subfloat{\includegraphics[width=\linewidth]{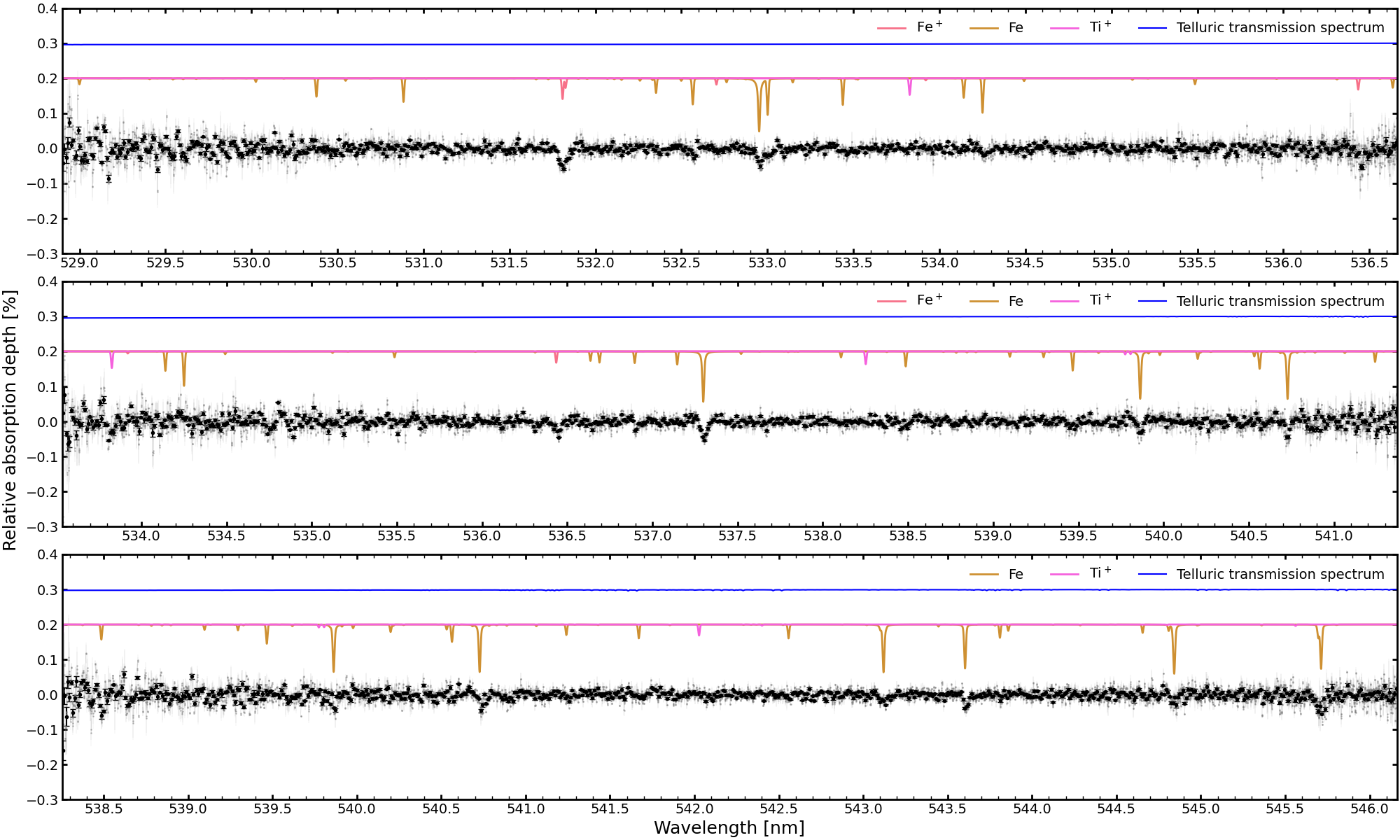}}\\[1em]
    Fig.\,\ref{fig:atlas_0_blue} continued.
\end{sidewaysfigure}

\begin{sidewaysfigure}
    \centering
    \ContinuedFloat
    \subfloat{\includegraphics[width=\linewidth]{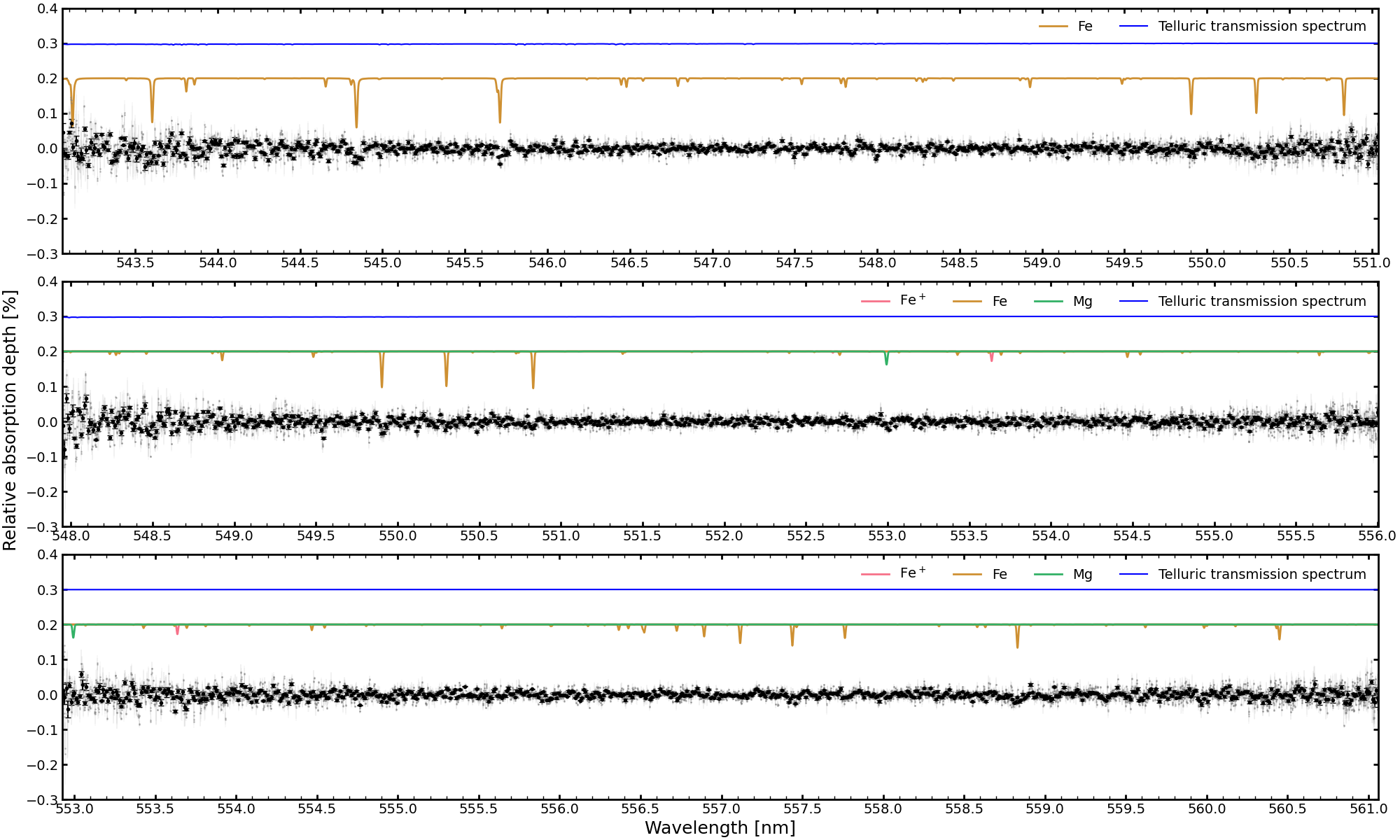}}\\[1em]
    Fig.\,\ref{fig:atlas_0_blue} continued.
\end{sidewaysfigure}

\begin{sidewaysfigure}
    \centering
    \ContinuedFloat
    \subfloat{\includegraphics[width=\linewidth]{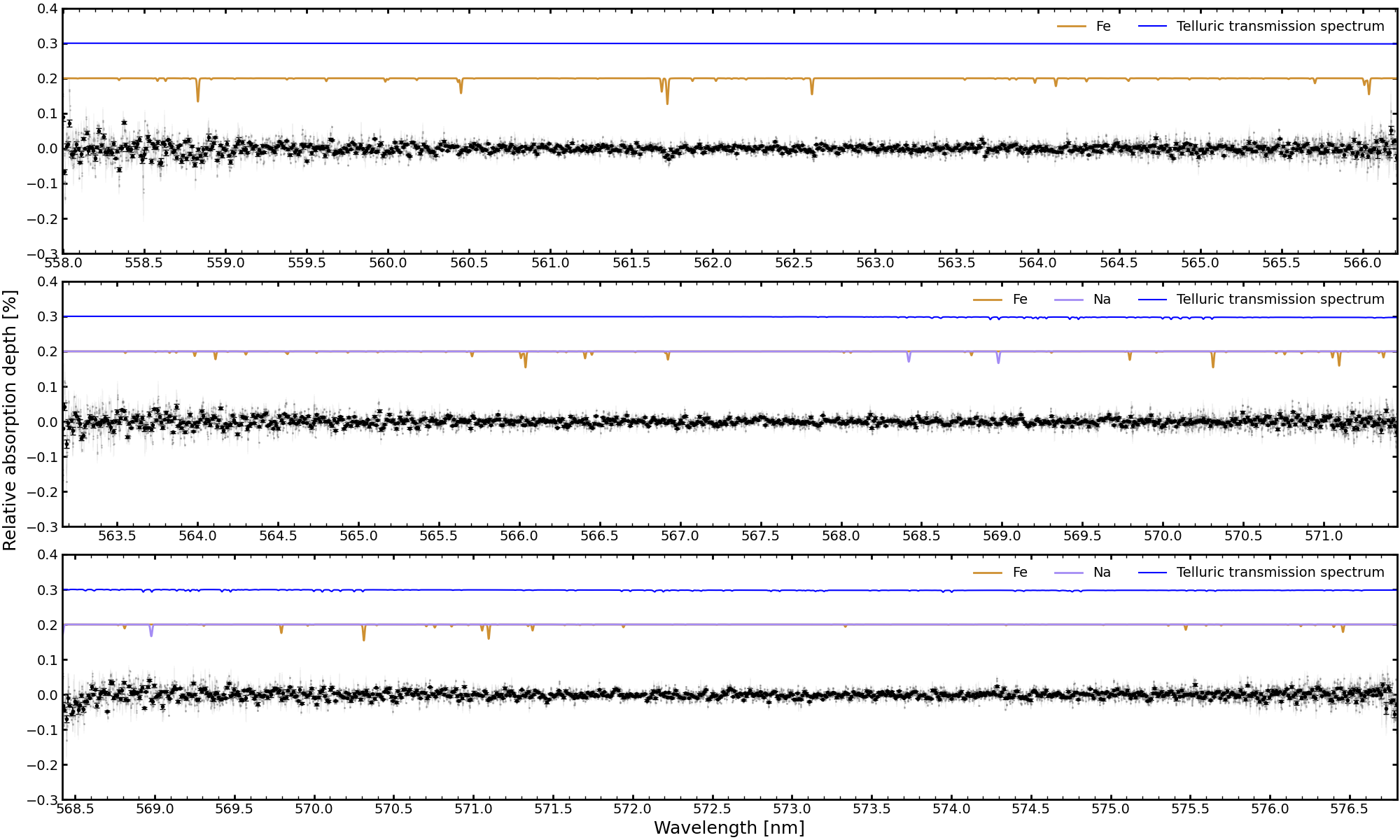}}\\[1em]
    Fig.\,\ref{fig:atlas_0_blue} continued.
\end{sidewaysfigure}

\begin{sidewaysfigure}
    \centering
    \ContinuedFloat
    \subfloat{\includegraphics[width=\linewidth]{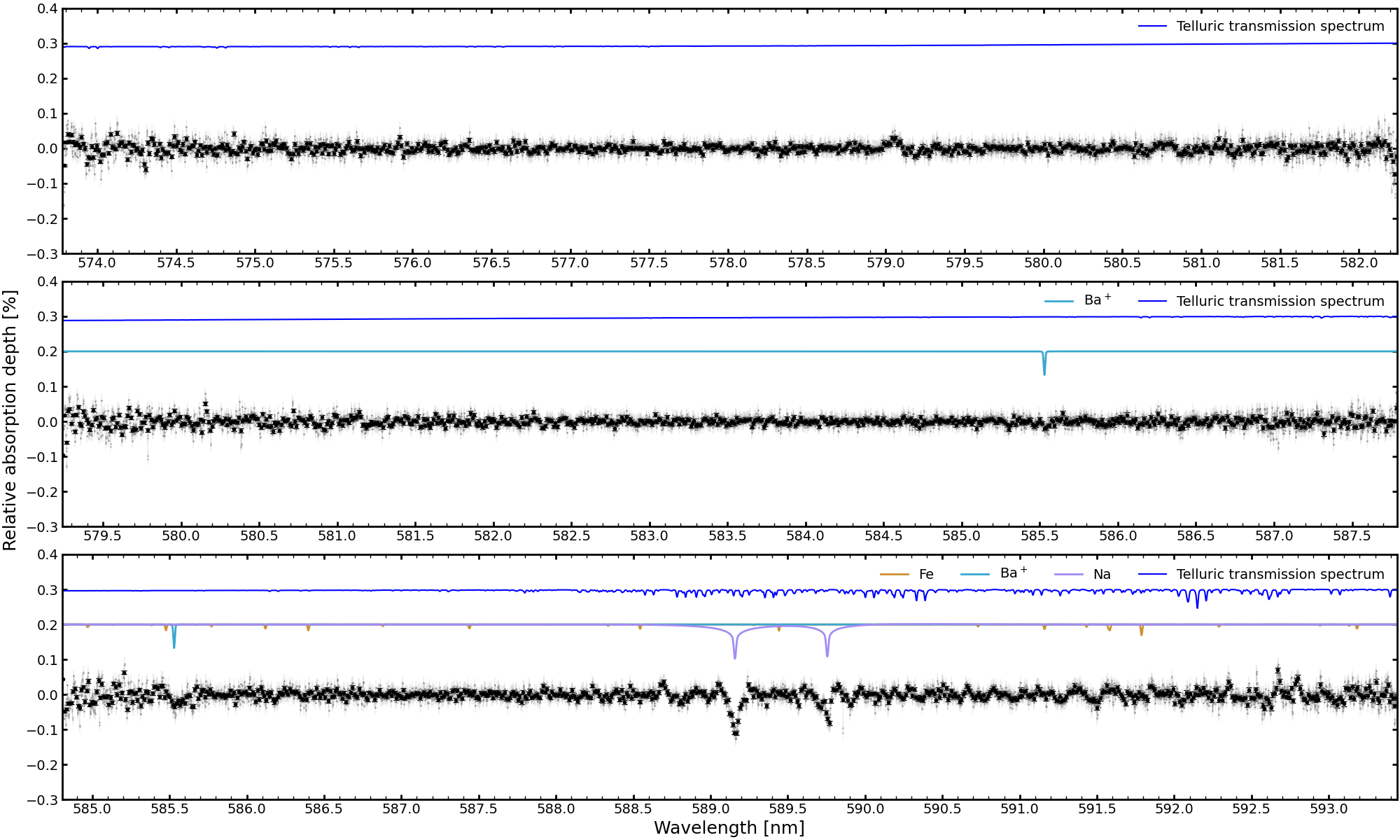}}\\[1em]
    Fig.\,\ref{fig:atlas_0_blue} continued.
\end{sidewaysfigure}

\begin{sidewaysfigure}
    \centering
    \ContinuedFloat
    \subfloat{\includegraphics[width=\linewidth]{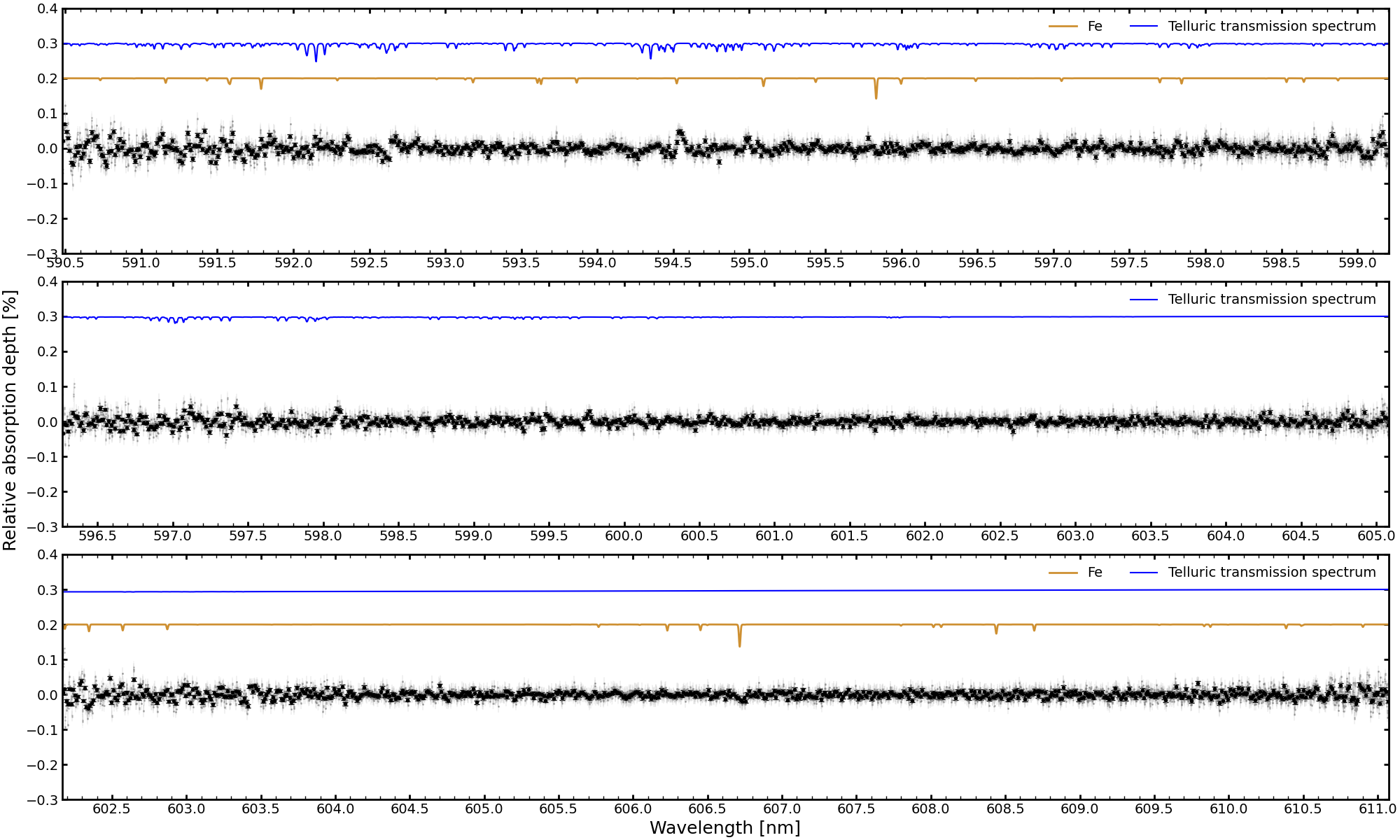}}\\[1em]
    Fig.\,\ref{fig:atlas_0_blue} continued.
\end{sidewaysfigure}

\begin{sidewaysfigure}
    \centering
    \ContinuedFloat
    \subfloat{\includegraphics[width=\linewidth]{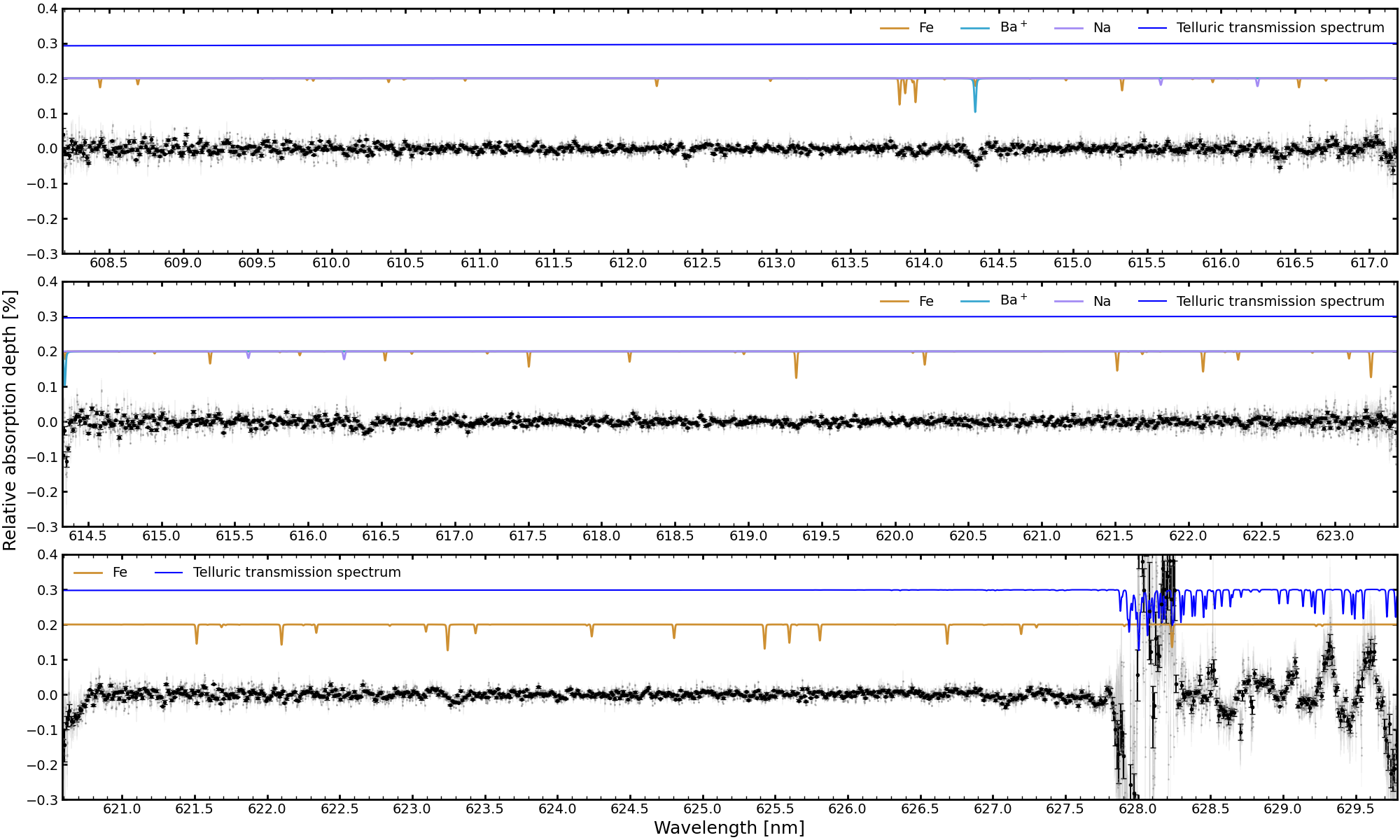}}\\[1em]
    Fig.\,\ref{fig:atlas_0_blue} continued.
\end{sidewaysfigure}

\begin{sidewaysfigure}
    \centering
    \ContinuedFloat
    \subfloat{\includegraphics[width=\linewidth]{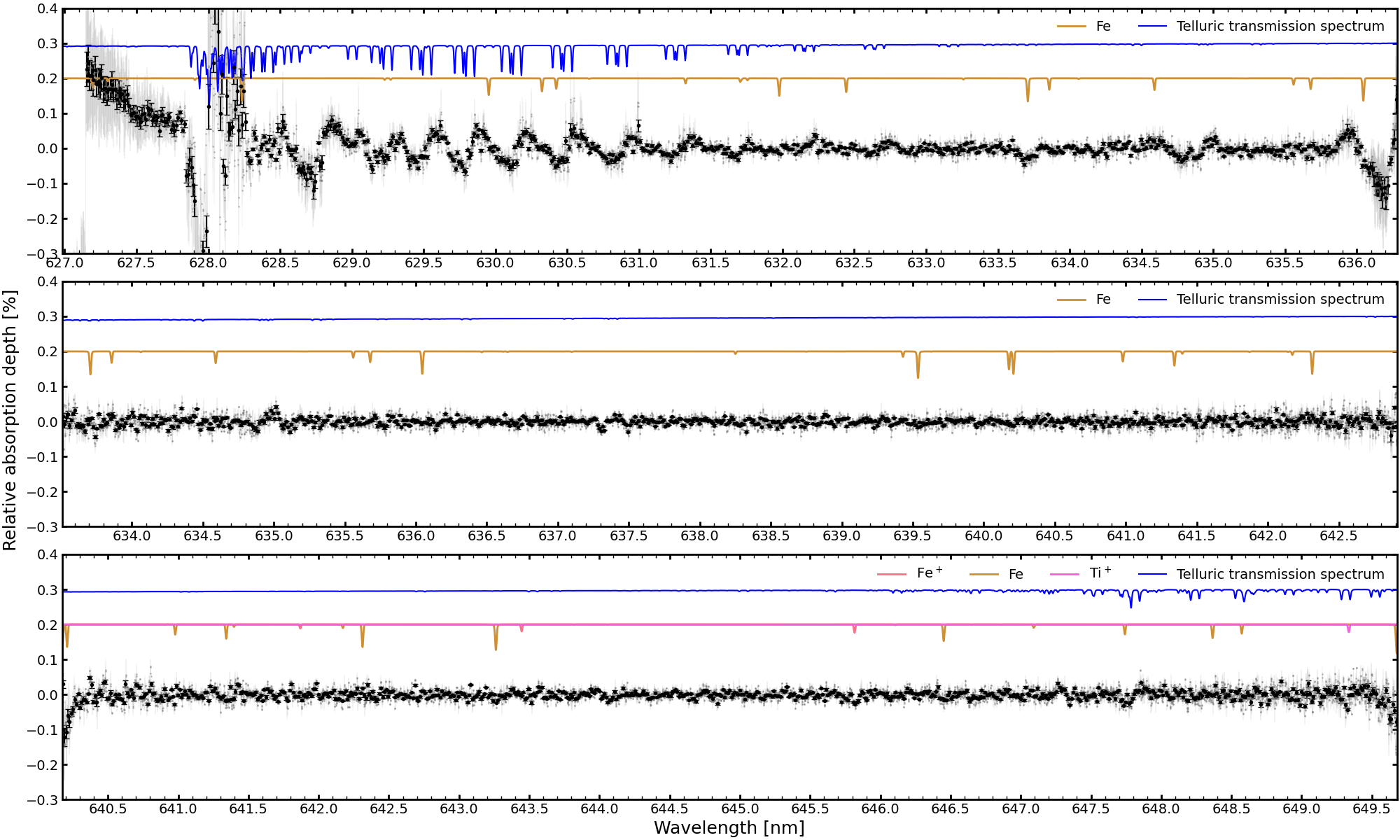}}\\[1em]
    Fig.\,\ref{fig:atlas_0_blue} continued.
\end{sidewaysfigure}

\begin{sidewaysfigure}
    \centering
    \ContinuedFloat
    \subfloat{\includegraphics[width=\linewidth]{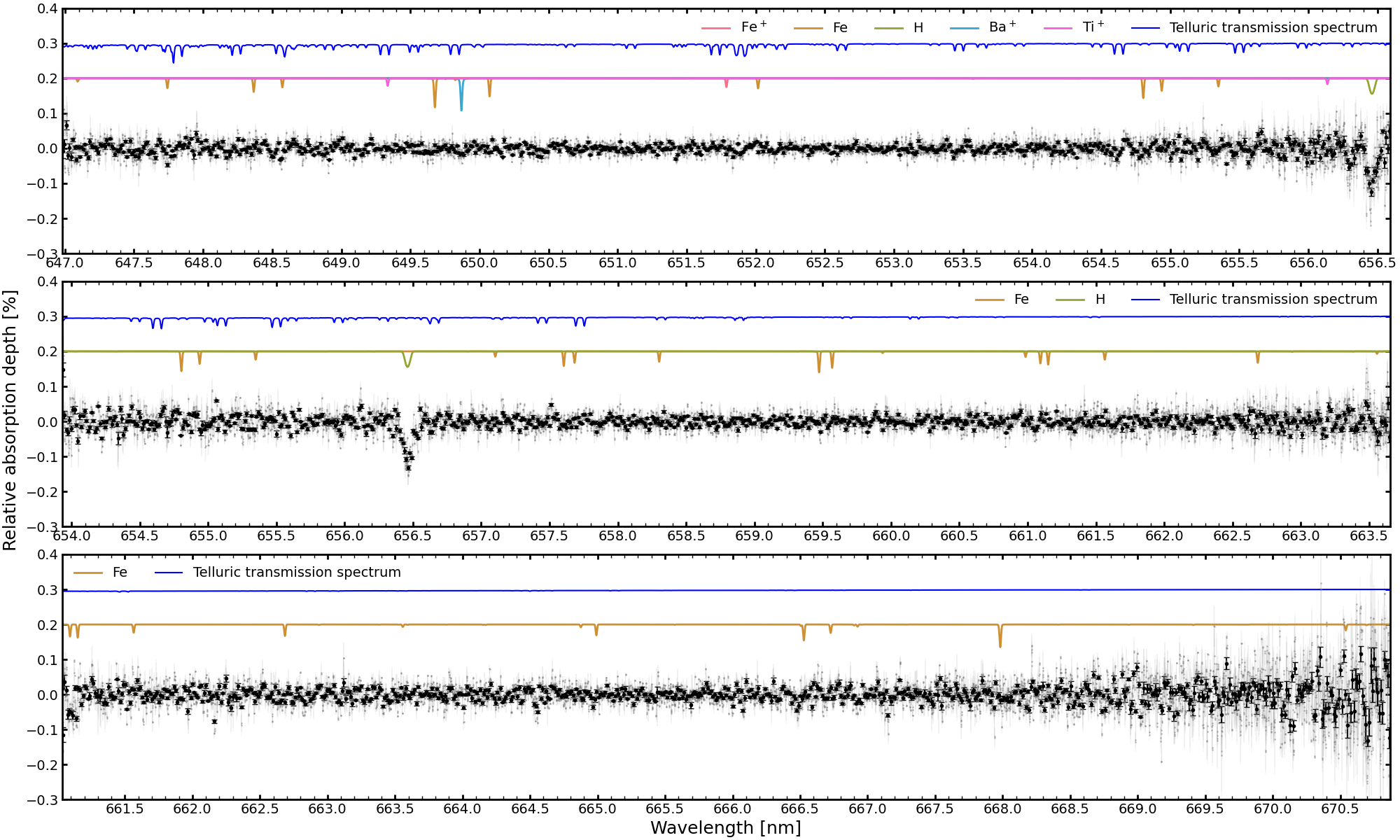}}\\[1em]
    Fig.\,\ref{fig:atlas_0_blue} continued.
\end{sidewaysfigure}

\begin{sidewaysfigure}
    \centering
    \subfloat{\includegraphics[width=\linewidth]{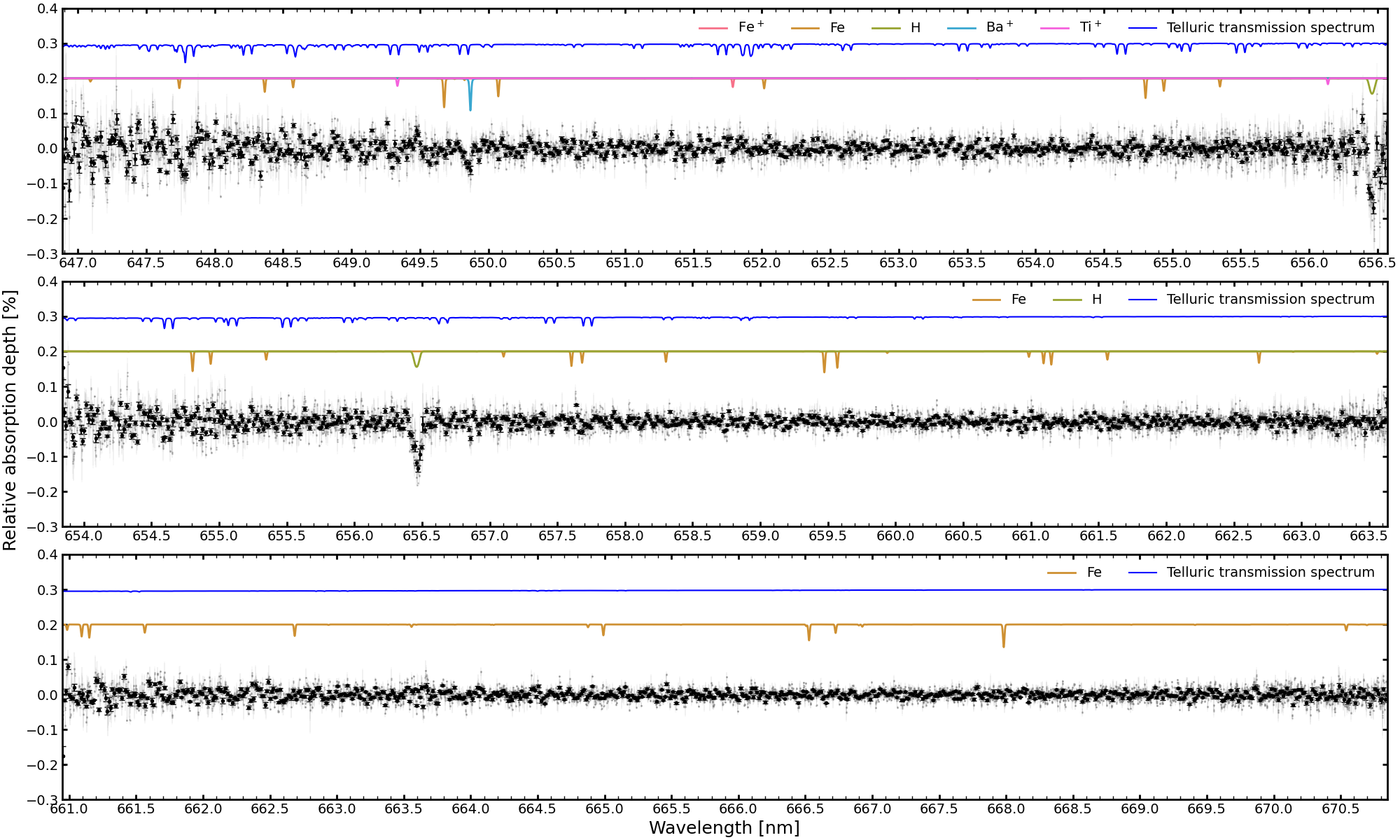}}
    \caption{Same as Fig.\,\ref{fig:atlas_0_blue} but for the red arm of MAROON-X. The spectra of orders 13 and 14  are not shown because of heavy telluric contamination by the \ch{O2} band.}
    \label{fig:atlas_0_red}
\end{sidewaysfigure}

\begin{sidewaysfigure}
    \centering
    \ContinuedFloat
    \subfloat{\includegraphics[width=\linewidth]{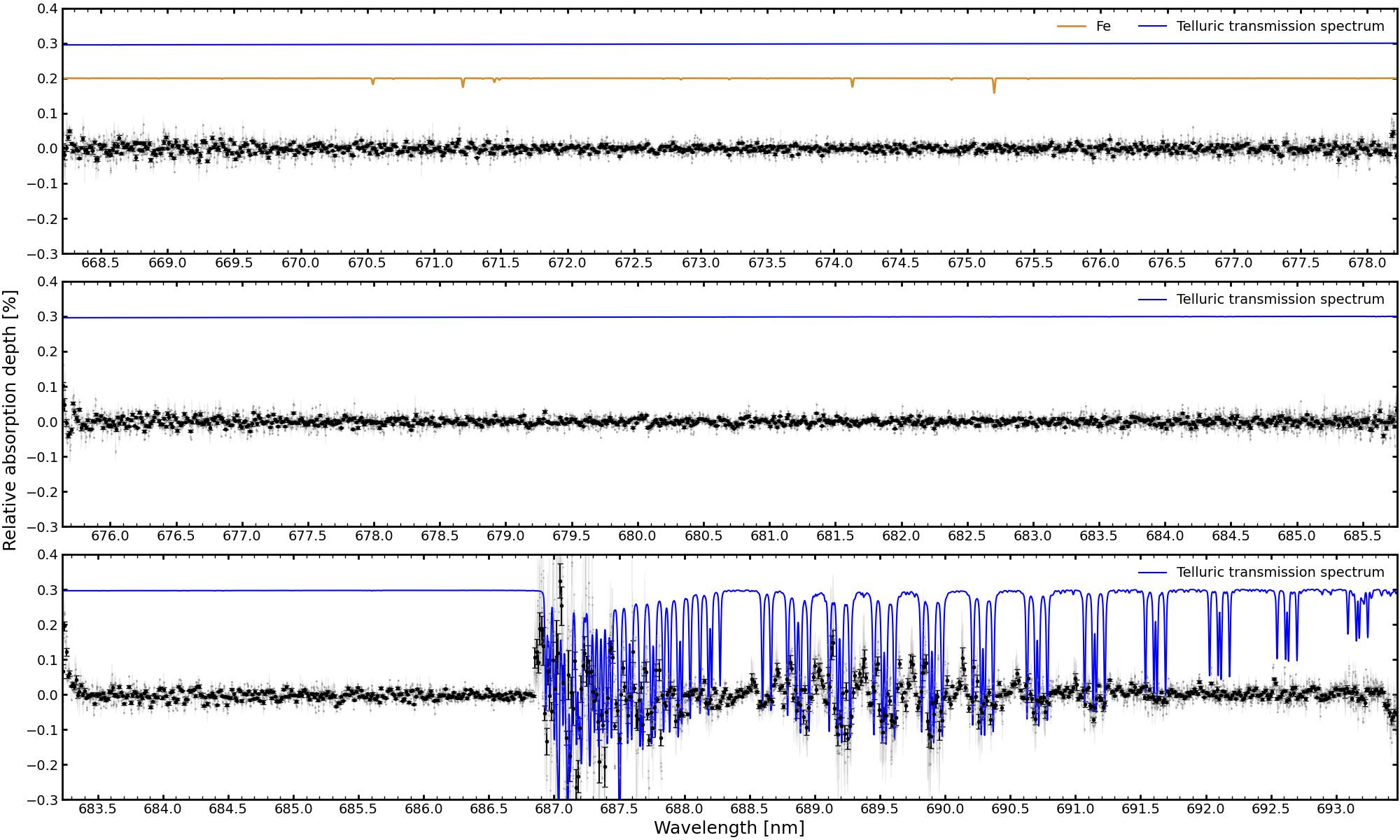}}\\[1em]
    Fig.\,\ref{fig:atlas_0_red} continued.
\end{sidewaysfigure}

\begin{sidewaysfigure}
    \centering
    \ContinuedFloat
    \subfloat{\includegraphics[width=\linewidth]{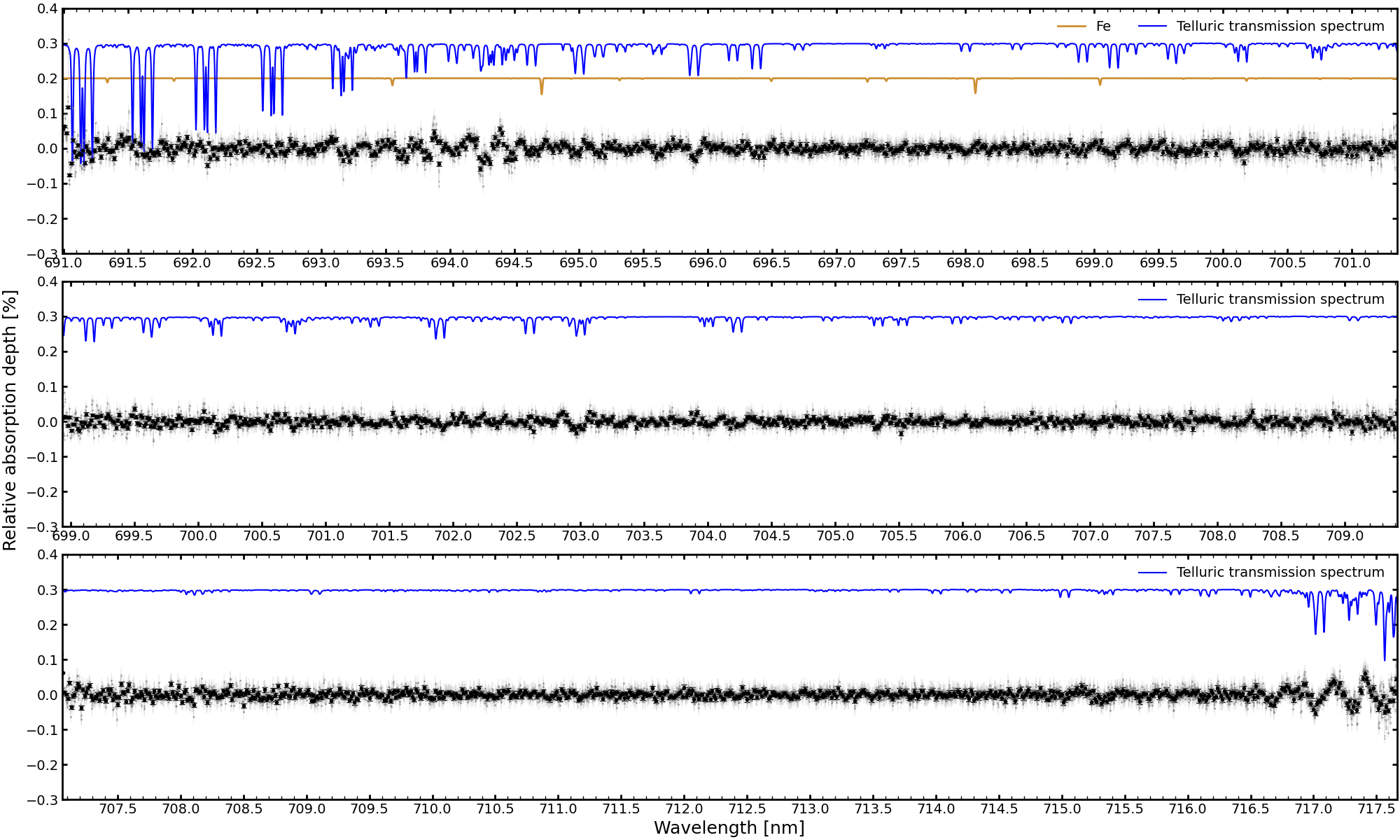}}\\[1em]
    Fig.\,\ref{fig:atlas_0_red} continued.
\end{sidewaysfigure}

\begin{sidewaysfigure}
    \centering
    \ContinuedFloat
    \subfloat{\includegraphics[width=\linewidth]{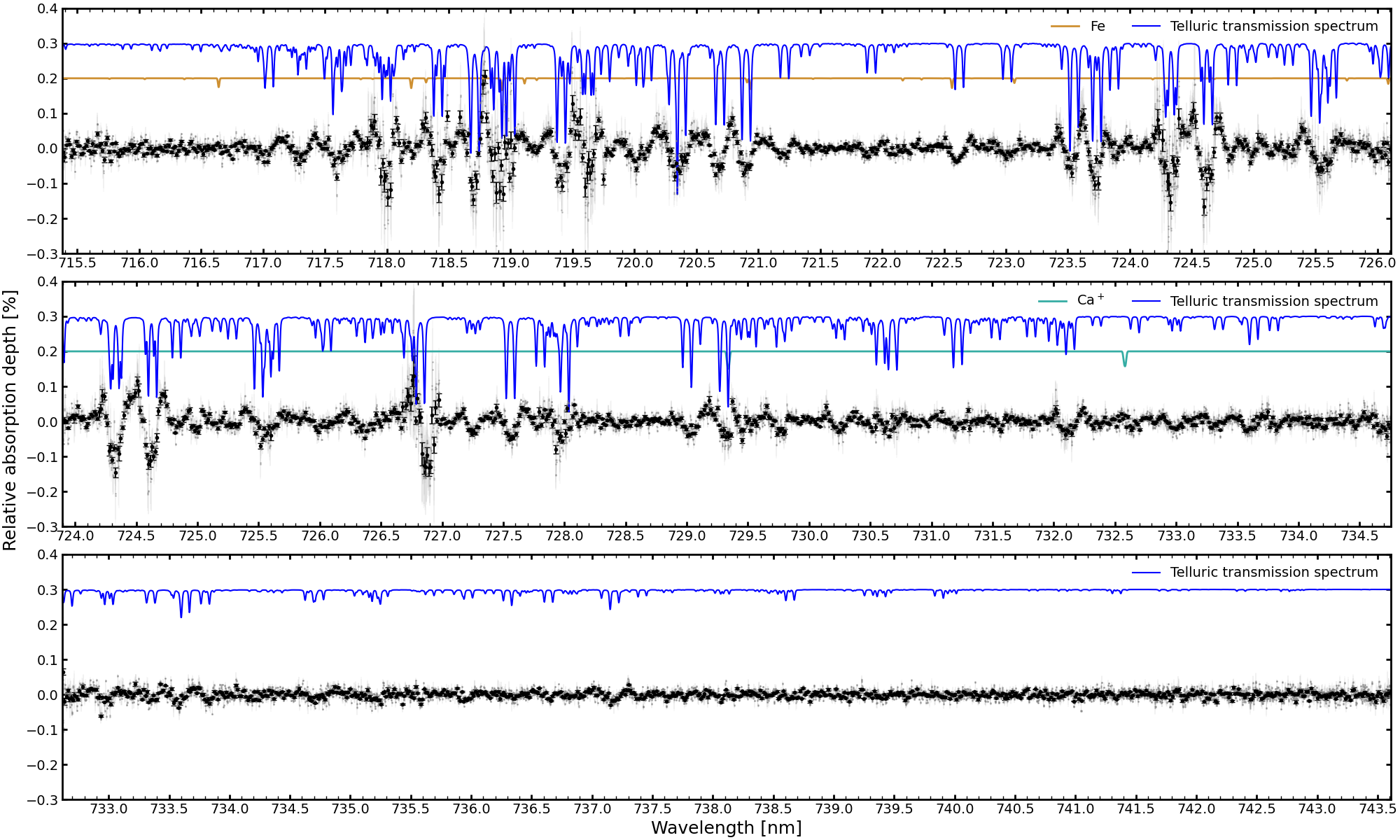}}\\[1em]
    Fig.\,\ref{fig:atlas_0_red} continued.
\end{sidewaysfigure}

\begin{sidewaysfigure}
    \centering
    \ContinuedFloat
    \subfloat{\includegraphics[width=\linewidth]{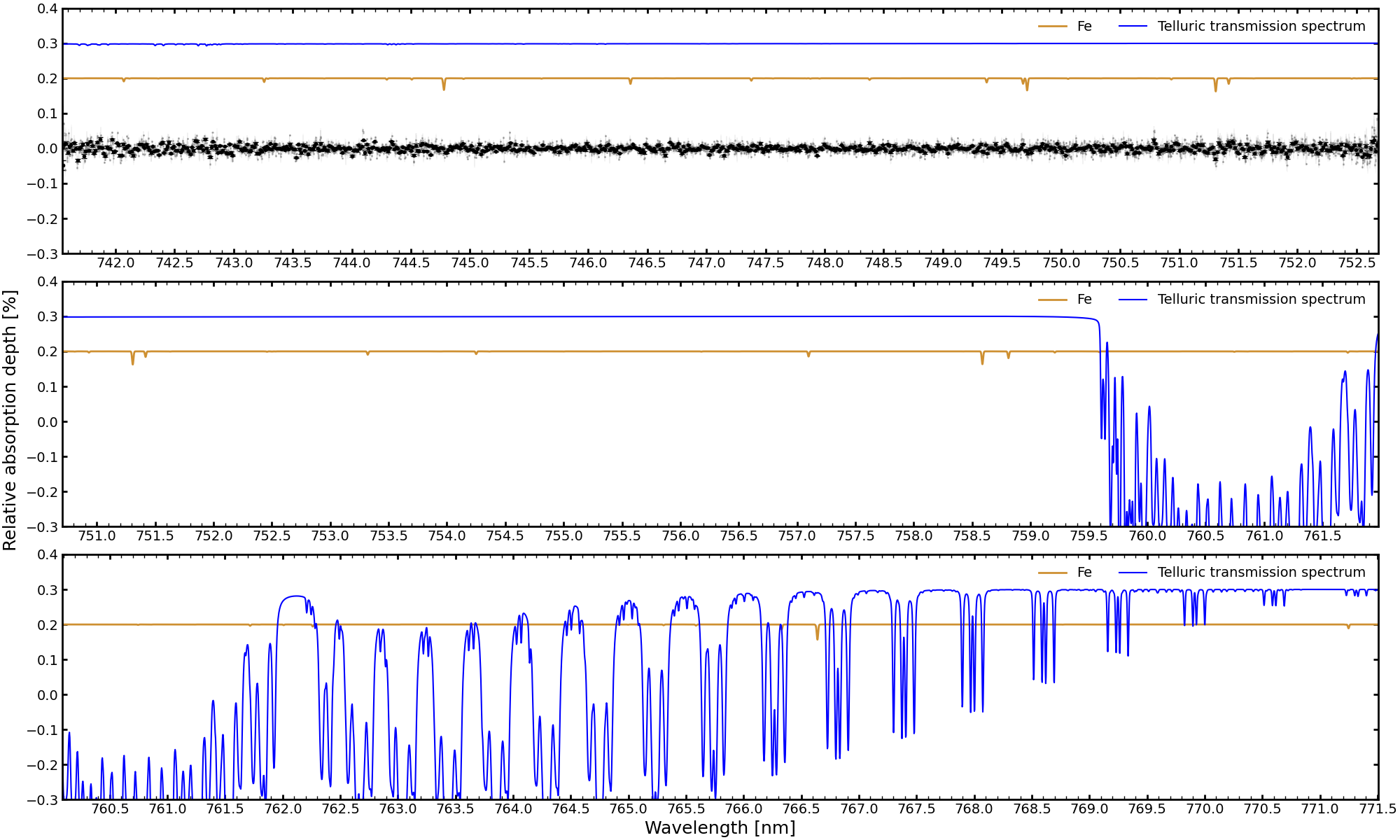}}\\[1em]
    Fig.\,\ref{fig:atlas_0_red} continued.
\end{sidewaysfigure}

\begin{sidewaysfigure}
    \centering
    \ContinuedFloat
    \subfloat{\includegraphics[width=\linewidth]{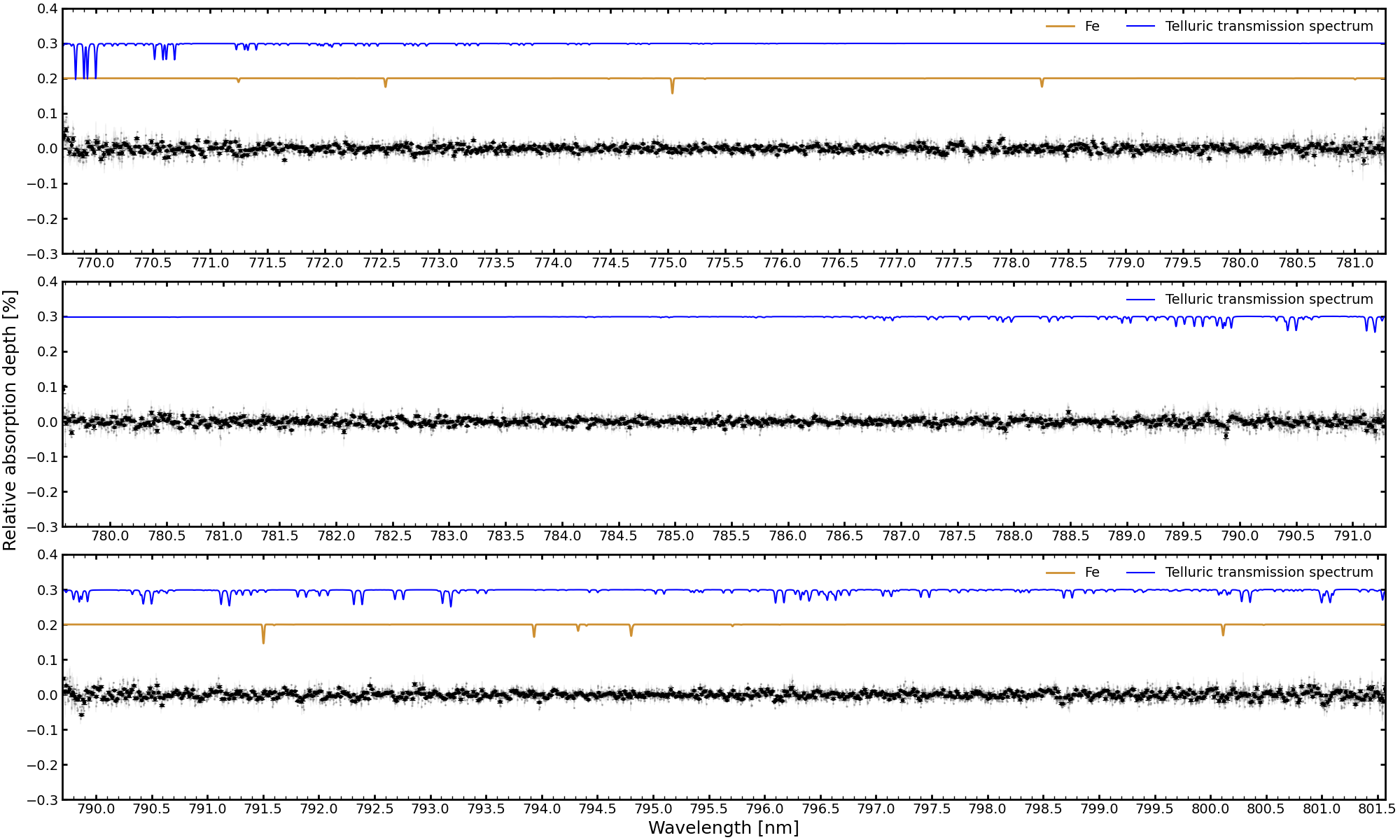}}\\[1em]
    Fig.\,\ref{fig:atlas_0_red} continued.
\end{sidewaysfigure}

\begin{sidewaysfigure}
    \centering
    \ContinuedFloat
    \subfloat{\includegraphics[width=\linewidth]{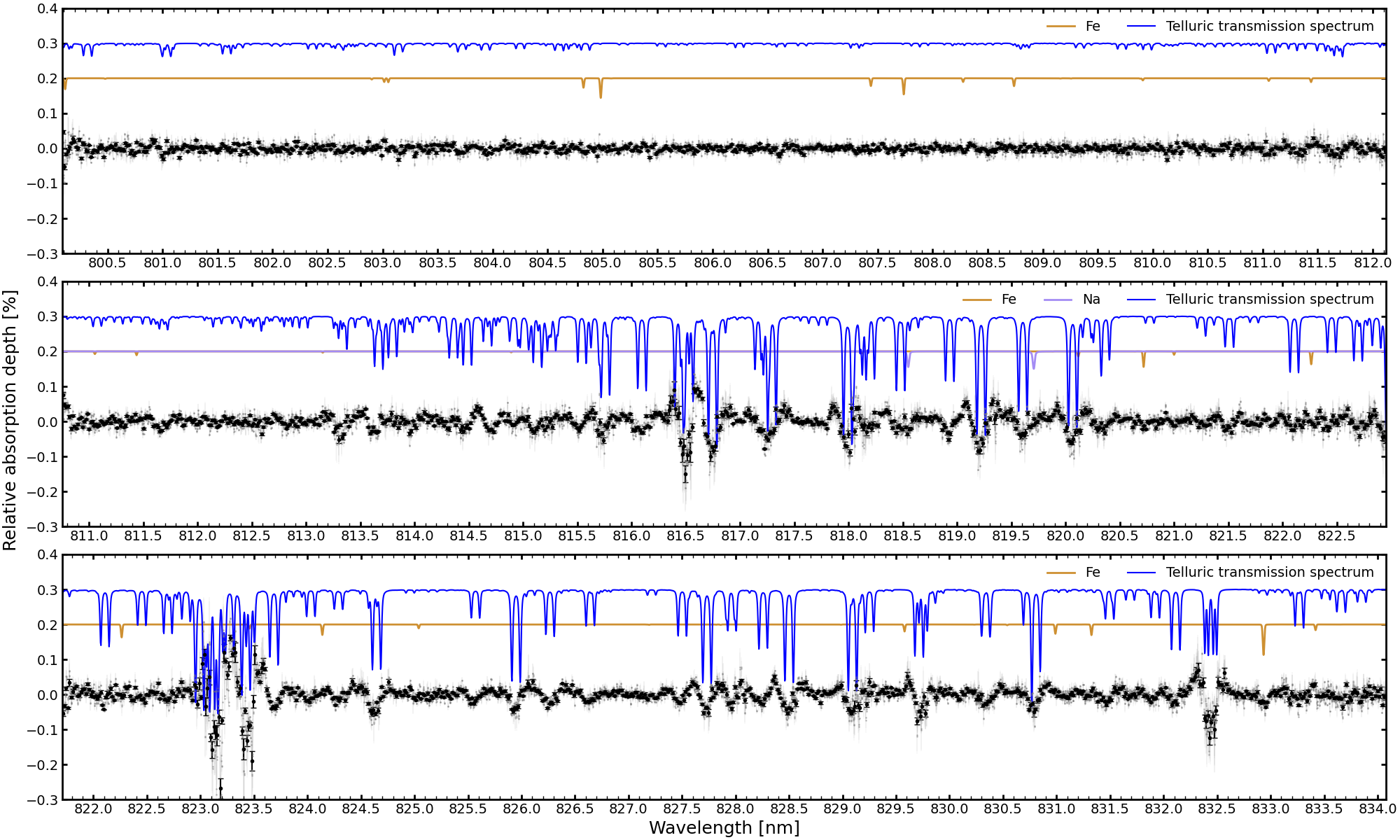}}\\[1em]
    Fig.\,\ref{fig:atlas_0_red} continued.
\end{sidewaysfigure}

\begin{sidewaysfigure}
    \centering
    \ContinuedFloat
    \subfloat{\includegraphics[width=\linewidth]{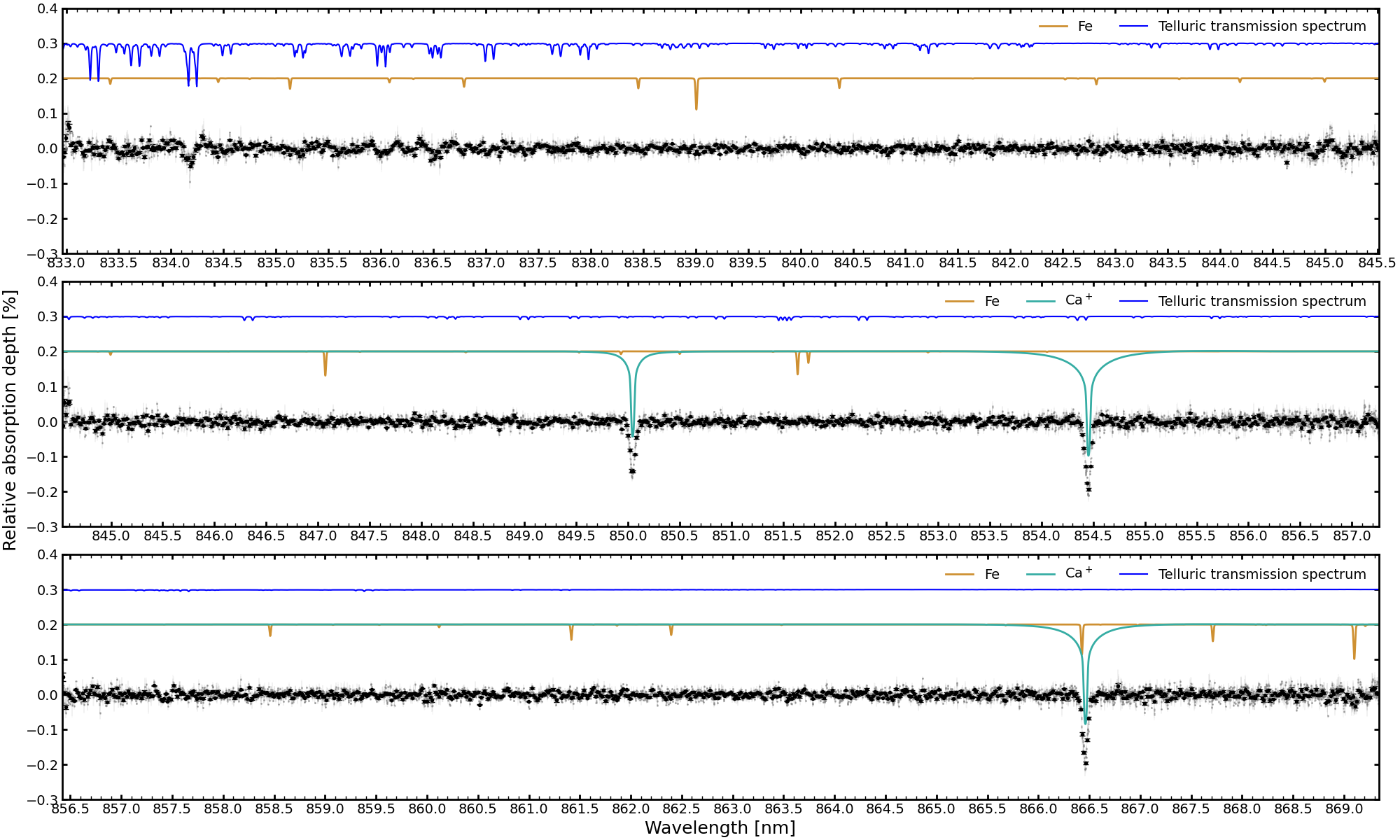}}\\[1em]
    Fig.\,\ref{fig:atlas_0_red} continued.
\end{sidewaysfigure}

\begin{sidewaysfigure}
    \centering
    \ContinuedFloat
    \subfloat{\includegraphics[width=\linewidth]{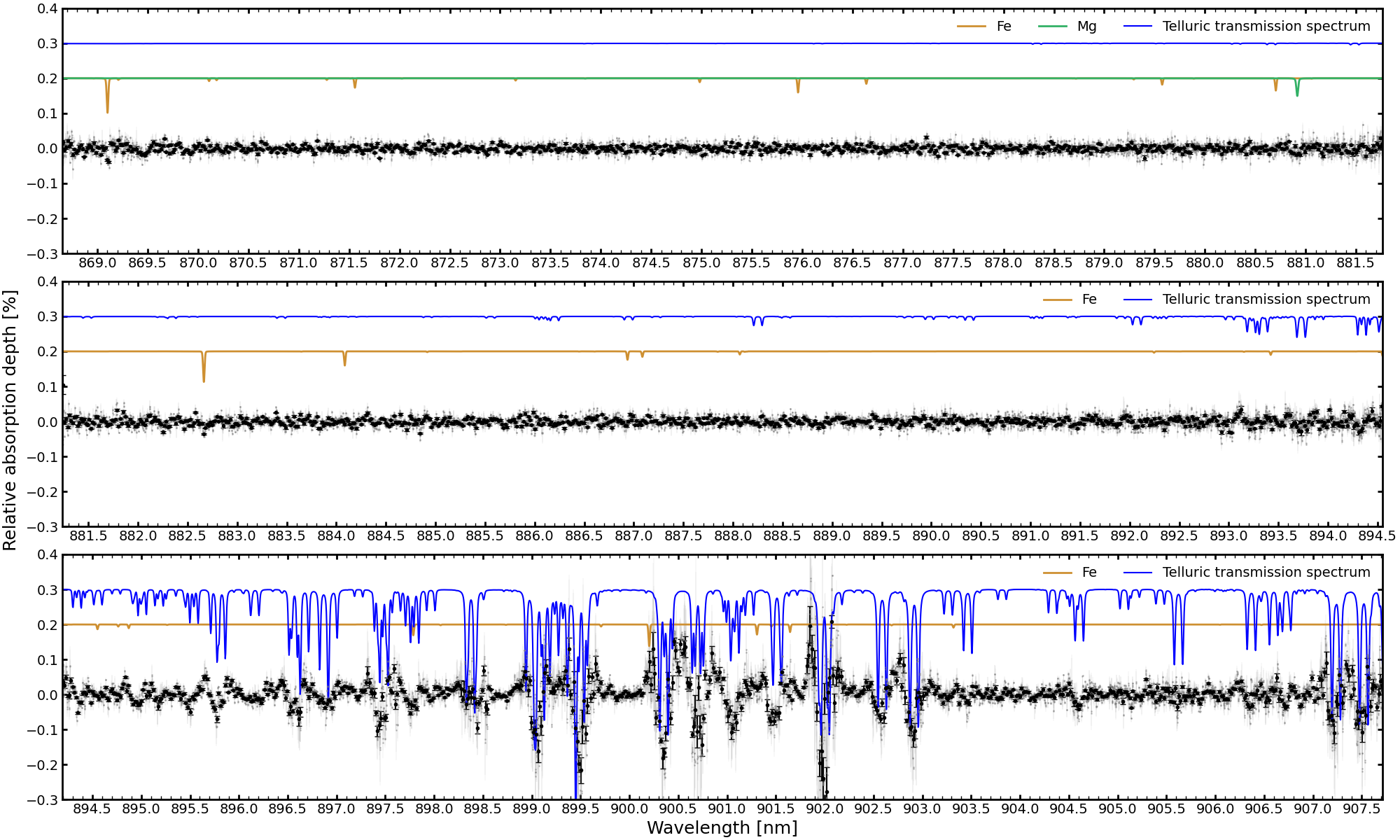}}\\[1em]
    Fig.\,\ref{fig:atlas_0_red} continued.
\end{sidewaysfigure}

\begin{sidewaysfigure}
    \centering
    \ContinuedFloat
    \subfloat{\includegraphics[width=\linewidth]{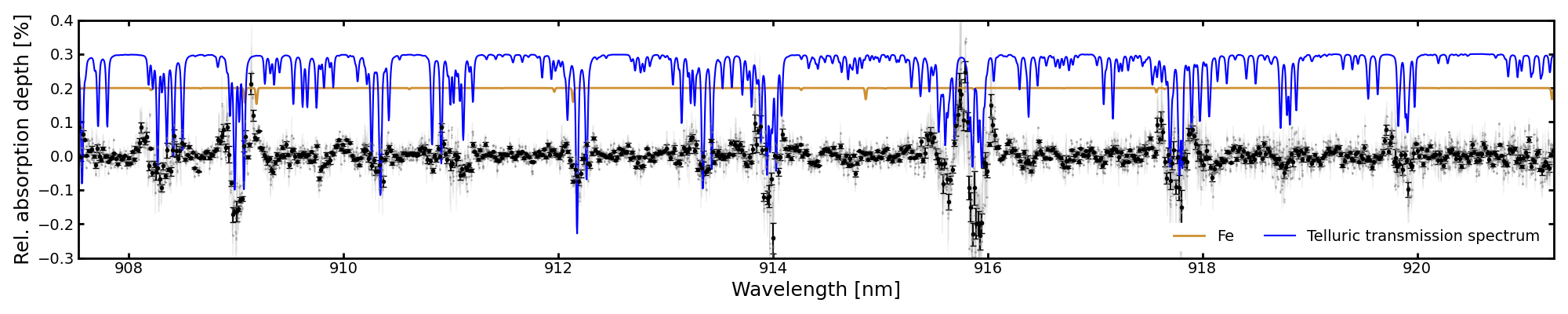}}\\[1em]
    Fig.\,\ref{fig:atlas_0_red} continued.
\end{sidewaysfigure}

\end{appendix}
\end{document}